\documentclass[a4paper,11pt]{article}
\usepackage{jheppub}

\usepackage{calc}
\usepackage{rotating}
\usepackage[english]{babel}
\usepackage{graphicx}
\usepackage{subfig}
\usepackage{float}
\usepackage{amsmath}
\usepackage{amssymb}
\usepackage{amsthm}
\usepackage{latexsym}
\usepackage{color}
\usepackage{slashed}
\usepackage{multirow}
\usepackage[table]{xcolor}
\usepackage{color, colortbl}
\usepackage{booktabs}
\usepackage{gensymb}
\usepackage{array}

\newcommand{\newc}{\newcommand*}

\newc{\met}{\slashed E_T}
\newc{\neut}[1]{\ensuremath{\widetilde{\chi}^0_{#1}}}
\newc{\chgo}[1]{\ensuremath{\widetilde{\chi}^{\pm}_{#1}}}
\newc{\mneut}[1]{\ensuremath{m_{\widetilde{\chi}^0_{#1}}}}
\newc{\mchgo}[1]{\ensuremath{m_{\widetilde{\chi}^{\pm}_{#1}}}}
\newc{\dmn}[1]{\ensuremath{\Delta m_{\widetilde{\chi}^0_{#1}}}}
\newc{\dmc}{\ensuremath{\Delta m_{\widetilde{\chi}^\pm}}}

\newc{\gev}{\ensuremath{\,\mathrm{GeV}}}
\newc{\tev}{\ensuremath{\,\mathrm{TeV}}}
\newc{\sne}[1]{\ensuremath{S^{{#1},\,{\rm eff}}_1}}
\newc{\snet}{\ensuremath{S^{\rm eff}_1}}
\newc{\scet}{\ensuremath{S^{\rm eff}_2}}
\newc{\omegan}{\ensuremath{\Omega_{\widetilde{\chi}^0_{1}}}}     
\newc{\omegadm}{\ensuremath{\Omega_{{\rm DM}}}}     

\newc{\pos}{H}
\newc{\scale}{1}

\subheader{
\textrm{\begin{flushright}
KIAS-P17074\\
\end{flushright}}}

\title{Closing in on the Wino LSP via trilepton searches at the LHC}

\author[a]{W. Abdallah,}
\author[b]{S. Khalil,}
\author[c]{S. Moretti}
\author[d,1]{and S. Munir\note{Corresponding author.}}

\affiliation[a]{Department of Mathematics, Faculty of Science,\\Cairo University, Giza, Egypt}
\affiliation[b]{Center for Fundamental Physics, Zewail City of Science and Technology,\\6 October City, Giza, Egypt} 
\affiliation[c]{School of Physics \& Astronomy, University of Southampton,\\Highfield, Southampton SO17 1BJ, UK}
\affiliation[d]{School of Physics, Korea Institute for Advanced Study,\\Seoul 130-722, Republic of Korea}

\emailAdd{awaleed@sci.cu.edu.eg}
\emailAdd{skhalil@zewailcity.edu.eg}
\emailAdd{s.moretti@soton.ac.uk}
\emailAdd{smunir@kias.re.kr}

\abstract{
The neutralino dark matter (DM) predicted by the Minimal Supersymmetric Standard Model (MSSM) has been probed in several search modes at the Large Hadron Collider (LHC), one of the leading ones among which is the trilepton plus missing transverse momentum channel. The experimental analysis of this mode has, however, been designed to probe mainly a bino-like DM, originating in the decays of a pair of next-to-lightest neutralino and lightest chargino, both of which are assumed to be wino-like. In this study, we analyse how this trilepton channel can be tuned for probing also the wino-like DM. We note that, while the mentioned standard production mode generally leads to a relatively poor sensitivity for the wino-like DM, there are regions in the MSSM parameter space where the net yield in the trilepton final state can be substantially enhanced at the LHC with $\sqrt{s}=14$\,TeV. This is achieved by taking into account also an alternative channel, pair-production of the wino-like DM itself in association with the heavier chargino, and optimisation of the kinematical cuts currently employed by the LHC collaborations. In particular, we find that the cut on the transverse mass of the third lepton highly suppresses both the signal channels and should therefore be discarded in this DM scenario. We perform a detailed detector-level study of some selected parameter space points that are consistent with the most important experimental constraints, including the recent ones from the direct and indirect DM detection facilities. Our analysis demonstrates the high complementarity of the two channels, with their combined significance reaching above 4$\sigma$ for a wino-like DM mass around 100\,GeV, with an integrated luminosity as low as 100\,fb$^{-1}$.}

\begin{document}
\maketitle
\section{Introduction}
\label{sec:intro}

The MSSM contains four neutralinos, \neut{1-4}, which are the mass eigenstates resulting from the mixing of the fermion components of the Higgs superfields, known as the higgsinos ($\widetilde{H}_d^0,\widetilde{H}_u^0$), with those of the gauge superfields, the gauginos ($\widetilde{B}^0,\widetilde{W}^0$). The lightest of these neutralinos, \neut{1}, is an important DM candidate when it is the Lightest Supersymmetric Particle (LSP) and $R$-parity is conserved. Whether or not it can {\it thermally} generate the observed DM relic density of the universe, \omegadm, as measured by the WMAP~\cite{Hinshaw:2012aka} and PLANCK~\cite{Planck:2015xua} telescopes, depends strongly on the interplay between its mass and composition. This composition is governed by the relative sizes of the soft Supersymmetry (SUSY)-breaking gaugino mass parameters, $M_1$ and $M_2$, and the Higgs-higgsino mass parameter, $\mu$, originating in the MSSM superpotential. The \neut{1} can be generally categorised as bino-, wino- or higgsino-like, if the splitting between the magnitudes of these parameters in the neutralino mass matrix is fairly large. Conversely, when all of them have values lying close to each other, the DM is a highly mixed state.  

A higgsino-like DM is theoretically motivated by the naturalness requirement of $|\mu|\sim 200$\,GeV~\cite{Brust:2011tb,Papucci:2011wy,Hall:2011aa,Feng:2012jfa,Cao:2012rz,Baer:2012up,Han:2013poa,Han:2013kga,Kowalska:2013ica}, although the \neut{1} mass obtained for such a low $|\mu|$ is not favoured by the \omegadm\ measurements~\cite{ArkaniHamed:2006mb}. A bino-like $\neut{1}$ with mass $\mathcal{O}$(10)\,GeV can, in contrast, not only be exactly consistent with the PLANCK measurement of \omegadm\ (see, e.g.,~\cite{Bergeron:2013lya}) but also explain the galactic centre $\gamma$-ray excess~\cite{Hooper:2010mq,Hooper:2011ti} observed by the Fermi Large Area Telescope (FermiLAT). Such a DM is also typically predicted by minimal Supergravity (mSUGRA)-inspired boundary conditions in the MSSM, which lead to $M_1 < M_2$ at the electroweak (EW) scale. A wino-like DM is motivated in, contrary to the mSUGRA scenario, the Anomaly Mediated SUSY-Breaking (AMSB)  scenario~\cite{Giudice:1998xp,Randall:1998uk,Bagger:1999rd}. However, the self-annihilation of a purely wino-like DM as well as its co-annihilation with the lighter of the two charginos, \chgo{1,2}, which is almost mass-degenerate with it and hence the Next-to-LSP (NLSP), produces a thermal DM abundance that is a few orders of magnitude below \omegadm, unless its mass is above 2\,TeV~\cite{Beneke:2016ync}. For lower masses, a sufficient bino component can raise its abundance somewhat by reducing its interaction strength~\cite{Chakraborti:2017dpu}. Alternatively, one can assume either its non-thermal production, for instance, through a moduli decay in the early universe~\cite{Moroi:1999zb,Acharya:2008bk}, or the contribution of other long-lived particles to the current total \omegadm~\cite{Zurek:2008qg} (see also~\cite{Roszkowski:2017nbc}, and references therein, for a recent overview of various DM candidates and their searches).

At the LHC, many searches have been devised for supersymmetric DM, for a variety of its possible production modes. It can be pair-produced electroweakly, and the signature thus probed comprises of missing transverse energy, $\met$, along with a jet or a gauge boson coming from Initial State Radiation (ISR)~\cite{Gunion:1999jr,Giudice:2010wb,Goodman:2010ku,Fox:2011pm,Han:2013usa,Schwaller:2013baa,Baer:2014cua,Anandakrishnan:2014exa,Barducci:2015ffa}. However, the yield of this search channel is typically very small, owing to the statistical limitations caused by its simple kinematics and the large SM backgrounds. Alternatively, the \neut{1}\ may also originate in the decays of the heavier neutralinos and/or charginos, in which case the final state consists of $\met$ and one or more charged leptons, $\ell\equiv e^\pm,\,\mu^\pm$~\cite{Giudice:2010wb,Rolbiecki:2012gn,Han:2014kaa,Bramante:2014dza,Baer:2014kya,Han:2014xoa,Han:2015lma}. In particular, two Opposite Sign, Same Flavour (OSSF) leptons coming from the $\neut{2}\to \neut{1} Z^{(*)}$ decay provide important kinematical handles in signal selection, which can lead to an enhanced sensitivity. The most promising one among multilepton search channels though is the trilepton (two of which constitute the OSSF pair, while the third one is assumed to result from the decay of a \chgo{1}) plus $\met$ (i.e., 3$\ell + \met$) channel, since the kinematic distributions of such events can be rather distinct compared to the SM backgrounds. These searches have therefore long been considered important probes of DM, both at the Tevatron~\cite{Abachi:1995ek,Abe:1996ex,Baer:1999bq,Chertok:2008zz} and the LHC~\cite{Aad:2014nua,Khachatryan:2014qwa,ATLAS:2016uwq,CMS:2017fdz,ATLAS:2017uun,CMS:2017sqn}. 

In the case of the higgsino-like DM, since  \neut{2} (as well as \chgo{1}) typically lies very close in mass to \neut{1}, while the remaining two electroweakinos (EWinos) have much larger masses, the two OSSF leptons get softer as the \neut{1} - \neut{2} mass gap becomes smaller. While an accompanying hard ISR can be utilised to trigger on, the yield gets enhanced  only for $\dmn{2} \lesssim 50$\,GeV, with $\dmn{i} \equiv \mneut{i} - \mneut{1}$~\cite{Gori:2013ala,vanBeekveld:2016hbo}. Thus, the event selection in the 3$\ell + \met$ searches has so far been optimised to favour a bino-like $\neut{1}$. In addition to the possibility of a large enough \dmn{i}\ for the outgoing leptons to be detectably hard, this has some other experimental underpinnings also, besides the theoretical motivations for a bino-like DM noted above. Such a DM, unlike the wino- or higgsino-like one, is not directly affected by the lower bound on the \chgo{1}\ mass ($\sim 100$\,GeV) from the Large Electron Positron (LEP) collider and can thus have a mass comparatively much smaller. Such a low (assumed) DM mass helps reduce the phase-space suppression of the production cross section of the decaying heavier neutralino and chargino. Furthermore, since the two parent particles both have nearly the same (wino-like, higgsino-like or wino-higgsino) composition, their masses are typically identical, which cuts down the number of variables in the kinematic selection of events.  

The aim of this article is to show that the 3$\ell + \met$ searches can have a sizeable sensitivity also to a wino-like DM at the LHC with $\sqrt{s}=14$\,TeV. This is made possible by the contribution of two processes, which can complement each other substantially in the case of a wino-like DM, to this final state. For exploring this possibility, we first find regions with a wino-like DM in the parameter space of the MSSM by performing  numerical scans, after imposing universality conditions on some of the parameters that do not have a significant role to play in this context. The scanned points are tested against the most crucial and recent experimental constraints, and the successful ones are used to study some important characteristics of these regions. We finally perform a thorough signal-to-background analysis of some benchmark points (BPs) in order to calculate the combined statistical significance of the two signal channels. In doing so, we highlight which kinematical cuts currently established in the experimental analyses need to be optimised for an improved detectability of the wino-like DM. 
One cut of particular relevance here is that on the transverse mass, which is imposed generally to minimise the $W^\pm Z$ background. However, in the DM scenario of our interest, the third lepton originates from a highly off-shell $W^{\pm *}$ and is thus very soft. Therefore, the transverse mass cut suppresses the signal processes much more strongly than the backgrounds, and dropping it results in an enhanced statistical significance.

The plan of of the paper is the following. In the next section, we briefly discuss the neutralino and chargino sectors of the MSSM as well as the two complementary signal processes of our interest. In section~\ref{sec:wino-DM} we dwell on the parameter space scans and describe in detail the properties of the points with wino-like DM obtained from these scans. In section~\ref{sec:analysis} we explain the specifics of our detector-level analysis as well as its results. We present our conclusions in section~\ref{sec:concl}.

\section{\label{sec:model} DM in the MSSM}

\subsection{\label{subsec:masses} Neutralino and chargino masses}

The mass matrix for the four neutralinos is given, in the $\widetilde{\psi}^0_j = (-i\widetilde B^0$,\,$-i\widetilde W^0$,\,$\widetilde H_u^0$,\,$\widetilde H_d^0$) basis, by
\begin{eqnarray}
{\cal M}_{\neut{}} =
\begin{pmatrix}
M_1 	& 0 	& -m_Z s_W c_\beta 	& m_Z s_W s_\beta	\\
 0	& M_2 	& m_Z c_W c_\beta 	& -m_Z c_W s_\beta 	\\
 -m_Z s_W c_\beta &  m_Z c_W c_\beta	& 0	&   -\mu	 \\
 m_Z s_W s_\beta	&  -m_Z c_W s_\beta 	&   -\mu	&   0	\\
\end{pmatrix}\,,
\label{eq:massmatrix}
\end{eqnarray}
where $s_W$ and $c_W$ are the sine and cosine of the weak mixing angle $\theta_W$, $m_Z$ is the mass of the $Z$ boson and $s_\beta$ and $c_\beta$ are short for $\sin\beta$ and $\cos\beta$, respectively, where $\beta$ is defined through $\tan\beta=v_u/v_d$, with $v_u$ and $v_d$ being the vacuum expectation values of the $\phi_u$ and $\phi_d$ Higgs doublets, respectively. The symmetric mass matrix in eq.~(\ref{eq:massmatrix}) can be diagonalised by a unitary matrix $N$ to give the diagonal matrix 
$D=\text{diag}(m_{\neut{i}}) = N^* {\cal M}_{\widetilde{\chi}^0} N^\dag$,
for $i=1 - 4$.\footnote{Note that in this study we assume all the parameters in ${\mathcal M}_{\neut{}}$ to be real. Thus,  the mixing matrix $N$ is an orthogonal matrix, so that $D = N {\cal M}_{\neut{}} N^T$. However, we use here the most general notation that allows for complex parameters in the mass matrices.}
The neutralino mass eigenstates are then given by $\neut{i} = N_{ij}^* \widetilde\psi^0_j$.
 
The eigenvalues of $N_{ij} $  can be positive or negative, but are not ordered in mass after performing the diagonalisation. They are reordered so that $\neut{1}$ is the
lightest eigenvalue, which is typically also the LSP and thus our DM candidate, and can be expressed as the linear combination
\begin{equation}
 \neut{1} = |N_{11}|^2 \widetilde B^0 + |N_{12}|^2 \widetilde W_3^0 + |N_{13}|^2 \widetilde H_d^0 + |N_{14}|^2 \widetilde H_u^0\,.
\end{equation}
It is clear that the sizes of $M_1$, $M_2$ and $\mu$, i.e., the bino, wino and higgsino mass parameters, respectively, describe the composition of the LSP. For example, if $M_2 \ll M_1,\,\mu$, then the LSP has a mass $m_{\tilde{\chi}^0_1} \simeq M_2$ and is thus referred to as `wino-like'. For convenience, we define 
\begin{equation}
  Z_B \equiv |N_{11}|^2\,,~Z_W \equiv |N_{12}|^2\,,~{\rm and}~Z_H \equiv |N_{13}|^2 + |N_{14}|^2,
\end{equation}
and, in the following, limit ourselves only to a `wino-dominated' DM candidate, i.e., to \neut{1} with $Z_W > \max(Z_B,\,Z_H)$. \\

The charged higgsinos ($\widetilde H_u^+$ and $\widetilde H_d^-$) and winos ($\widetilde{W}^+$ and $\widetilde{W}^-$) also mix to form the chargino eigenstates, $\widetilde{\chi}^\pm_a~ (a=1,2)$. The mass matrix for the charginos is given by
\begin{eqnarray}
{\cal M}_{\widetilde{\chi}^\pm} =
\begin{pmatrix}
M_2 	& \sqrt{2}m_W \sin\beta	 	\\
\sqrt{2}m_W \cos\beta	& \mu 	\\
 \end{pmatrix}\,,
\end{eqnarray}
where $m_W$ is the mass of the $W^\pm$ boson. By rotating this mass matrix using two unitary $2\times2$ matrices $U$ and $V$, the mass eigenvalues of the two physical charginos are obtained as
\begin{eqnarray}
m^2_{\widetilde \chi^{\pm}_{1,2}}
& = & {1\over 2} 
\Bigl [ |M_2|^2 + |\mu|^2 + 2m_W^2 \mp
\sqrt{(|M_2|^2 + |\mu |^2 + 2 m_W^2 )^2 - 4 |  M_2\mu - m_W^2 \sin 2
\beta |^2 }
\Bigr ]\,.\nonumber \\
\end{eqnarray}
In the limit of $M_2 \ll M_1,\, \mu$, where the LSP is wino-like, the lightest chargino is also dominantly wino-like and is nearly mass-degenerate with the LSP, as noted in the Introduction. 

\begin{figure}[tbp]
\begin{center}
\begin{tabular}{cc}
\vspace*{-1cm}\includegraphics*[width=6cm]{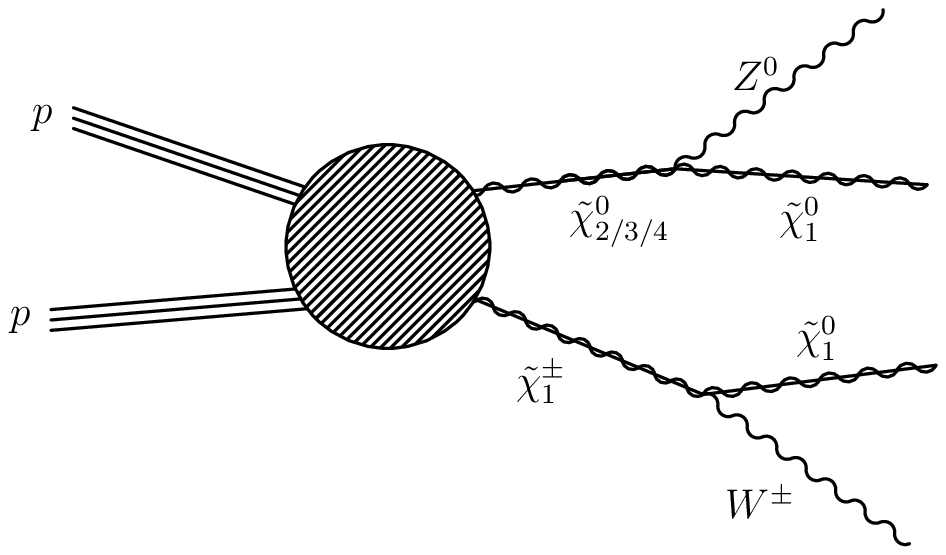} &
\hspace*{1cm}\includegraphics*[width=6cm]{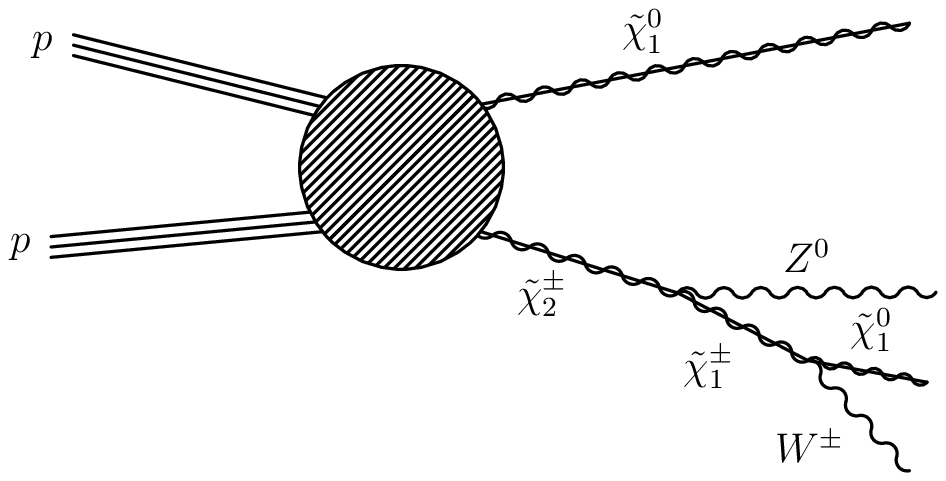} \\ 
 {\footnotesize (a)} & {\footnotesize (b)}
\end{tabular} 
\end{center}
\vspace*{-0.5cm}
\caption{\label{fig:processes} The two signal topologies, (a) $S_1$ and (b) $S_2$, that can contribute to the $3\ell +\met$ signature.} 
\end{figure}

\subsection{\label{subsec:trilep} Trilepton searches at the LHC}

At the LHC, the standard process that is assumed to give the 3$\ell + \met$ final state is the following one:
\begin{eqnarray*}
S^i_1\,&:&\,\,\, pp\to\widetilde \chi^0_{i} \, \widetilde \chi^{\pm}_{1}\to\widetilde \chi^0_{1} \, Z^{(*)} \, \widetilde \chi^0_{1} \, W^{\pm (*)}\to \widetilde \chi^0_{1} \, \ell^+\ell^- \,  \widetilde \chi^0_{1} \, \ell^\pm \, \nu_l\,,~{\rm with}~i=2,3,4\,.
\label{eq:S1}
\end{eqnarray*}
In the case of  wino-like DM, an alternative possibility, that has rarely been explored in literature, is the NLSP to be higgsino-like (implying $M_2 < \mu < M_1$). In such a scenario, not only are \neut{1} and \chgo{1} nearly mass-degenerate but also are $\neut{2,3}$ and $\chgo{2}$. This means that the mass difference between the two charginos, $\dmc \equiv \mchgo{2} - \mchgo{1}$, is very close to  $\dmn{2/3}$. This has some interesting phenomenological repercussions, with a crucial one being that, in certain regions of the MSSM parameter space, besides the standard $S_1^i$ channels noted above, the following process can contribute strongly to the 3$\ell + \met$ final state: 
\begin{eqnarray*}
S_2\,&:&\,\,\, pp\to\widetilde \chi^0_{1} \, \widetilde \chi^{\pm}_{2}\to\widetilde \chi^0_{1} \, \widetilde \chi^\pm_{1} \, Z^{(*)}\to \widetilde \chi^0_{1} \, \widetilde \chi^0_{1} \, W^{\pm (*)} \,  \ell^+\ell^- \to \widetilde \chi^0_{1} \, \widetilde \chi^0_{1} \, \ell^\pm \, \nu_l \, \ell^+\ell^-\,.
\end{eqnarray*}
The diagrammatic representation of the processes $S_1^i$ and $S_2$ is given in figure~\ref{fig:processes}. In the following sections we shall explore in depth these two processes and show that their complementarity can result in a large sensitivity to  wino-like DM at the LHC Run-II. 

\section{\label{sec:wino-DM} Parameter space scans and features}

\begin{table}[t!]
 \centering\begin{tabular}{cc}
\setlength\extrarowheight{5pt}
 \subfloat[]{
   \centering\begin{tabular}{|c|c|}
\hline 
Parameter & Scanned range  \\
\hline \hline
$M_1$\;(GeV)  & 10 -- 1000	\\
$M_2$\;(GeV)  & 90 -- 1000	\\
$\mu$\;(GeV) 	& ($\pm$)\,90 -- ($\pm$)\,1000 \\
$M_Q$\;(GeV) 	& 1000 -- 5000 \\
$M_L$\;(GeV) 	& 100 -- 3000 \\
$A_0$\; (GeV)  & $-7000$ -- $-500$\\
$\tan\beta$ 		& 2 -- 50 \\
$m_A$\;(GeV)  	& 125 -- 3000 \\
\hline
\end{tabular}
}
&
\setlength\extrarowheight{5pt}
\subfloat[]{
  \centering\begin{tabular}{|c|c|}
\hline
Observable & Measurement  \\
\hline \hline
${\rm BR}(B\to X_s \gamma) \times 10^{4} $   & $3.32\pm0.15$~\cite{Amhis:2016xyh}	\\ 
${\rm BR}(B_u\to \tau^\pm \nu_\tau) \times 10^{4} $ & $1.06\pm0.19$~\cite{Amhis:2016xyh}	\\ 
${\rm BR}(B_s \to \mu^+ \mu^-)\times 10^{9} $ & $3.0\pm 0.85$~\cite{Aaij:2017vad}\\ 
\hline
$\mu_{gg}$ & $1.14^{+0.19}_{-0.18}$~\cite{Khachatryan:2016vau}\\
$\mu_{ZZ}$ & $1.29^{+0.26}_{-0.23}$~\cite{Khachatryan:2016vau}\\
$\mu_{WW}$ & $1.09^{+0.18}_{-0.16}$~\cite{Khachatryan:2016vau}\\
$\mu_{\tau\tau}$  & $1.11^{+0.24}_{-0.22}$~\cite{Khachatryan:2016vau} \\
$\mu_{bb}$ & $0.70^{+0.29}_{-0.27}$~\cite{Khachatryan:2016vau} \\
\hline 
\end{tabular}
}
 \end{tabular}
\caption{\label{tab:para&cons} (a) MSSM parameters and their scanned ranges. (b) Experimental observables and their measured values, imposed as constraints on the scanned points.}
\end{table}

We first study some important general characteristics of the wino-dominated DM. For this purpose, we numerically scanned the parameter space of the phenomenological MSSM, wherein all the free parameters are input at the EW scale.
In order to reduce the number of free parameters, we imposed the following universality conditions on them:
\begin{gather}
  M_Q \equiv M_{Q_{1,2,3}} = M_{U_{1,2,3}} = M_{D_{1,2,3}}\,, \nonumber \\
  M_L \equiv M_{L_{1,2,3}} = M_{E_{1,2,3}}\,, \\
  A_0 \equiv A_{\tilde{t}} = A_{\tilde{b}} = A_{\tilde{\tau}}\,, \nonumber
\end{gather}
where $M_{Q_{1,2,3}},\,M_{U_{1,2,3}}$ and $M_{D_{1,2,3}}$ are the soft masses of the squarks, $M_{L_{1,2,3}}$ and $M_{E_{1,2,3}}$ those of the sleptons, and $A_{\tilde{t},\tilde{b},\tilde{\tau}}$ the soft trilinear couplings. These and the other free parameters were scanned in the ranges given in table~\ref{tab:para&cons}(a). In order to avoid any potential conflict of a scanned point with the null gluino searches at the LHC, we fixed $(m_{\tilde{g}}\sim )~ M_3 = 2$\,TeV. Note that we performed two separate scans corresponding to $\mu >0 $ and $\mu < 0$ each, since the DM-nucleon scattering cross section depends strongly on the sign of $\mu$ in addition to its magnitude, impacting the consistency of a given DM mass with direct detection limits. 

\subsection{\label{subsec:scan} Tools and methodology}

For each randomly generated point in the above parameter space, the mass spectra as well as the decay Branching Ratios (BRs) of the Higgs bosons and sparticles were calculated using the public code {\tt  
SPheno-v3.3.8}~\cite{Porod:2011nf,Porod:2003um}. For computing $\Omega_{\neut{1}} h^2$ and the spin-independent \neut{1}-proton scattering cross section, $\sigma^{\rm SI}_p$, we passed the output file generated by {\tt SPheno} for a given point to the code {\tt MicrOMEGAs-v4.3.1}~\cite{Belanger:2006is,Belanger:2013oya,Belanger:2014vza}. We then required each point to satisfy $\omegan h^2 \leq 0.131$, thus allowing a $+10\%$ error in the PLANCK measurement of $\omegadm = 0.119$~\cite{Planck:2015xua}. Moreover, in this study we identify the lightest CP-even Higgs boson, $h$, with the one observed at the LHC~\cite{Aad:2012tfa,Chatrchyan:2012ufa}. We therefore retained only points with $m_h=125\pm 2$\,GeV from the scans. We allow such flexibility in $m_h$, instead of enforcing the exact measurement ($125.09\pm 0.32$\,GeV~\cite{Aad:2015zhl}) on it, to accommodate hitherto unknown corrections from higher order calculations.

The points collected in the scans were further subjected to some other experimental constraints, which are listed in table~\ref{tab:para&cons}(b). The MSSM estimates of the $b$-physics observables shown in the table were obtained from {\tt SPheno} itself. The theoretical counterpart of the signal strength, $\mu_X$, of $h$ in the $X$ decay channel, can be defined (at the tree level) as
\begin{eqnarray}
\label{eq:Rxsct1}
R_X =  \frac{\sigma(pp \rightarrow
  h)}{\sigma(pp\rightarrow h_{\rm SM})}\times \frac{{\rm BR}(h
  \rightarrow X)}{{\rm BR}(h_{\rm SM} \rightarrow X)}\,,
\end{eqnarray}
where $h_{\rm SM}$ denotes the SM Higgs boson. We obtained these quantities for a given MSSM point from the public program {\tt HiggsSignals-v1.4.0}~\cite{Bechtle:2013xfa}. Note that, while the errors shown in the table are $1\sigma$, the points we consider allowed actually have model predictions lying within $2\sigma$ of the corresponding measured central values. Finally, the scanned points were also tested with {\tt HiggsBounds-v4.3.1}~\cite{Bechtle:2008jh,Bechtle:2011sb,Bechtle:2013wla}, and only the ones for which the heavier Higgs bosons $(H,\,A)$ were consistent with the exclusion bounds from Tevatron and LHC were used for subsequent analysis.

\subsection{\label{subsec:wino} Phenomenology of Wino-like DM}

Since the dominant contribution to  $\neut{i}\chgo{j}$ production comes from an $s$-channel $W^{\pm(*)}$, we begin our analysis by first calculating the phase-space independent effective quantities 
\begin{eqnarray}
S^{i,\,{\rm eff}}_1\,&:&\,\,\, g^2_{W \neut{i} \, \chgo{1}}\,\times\,{\rm BR}(\neut{i} \to  \neut{1} \, \mu^+\mu^-) \,\times\, {\rm BR}(\chgo{1} \to \neut{1}\,\mu^\pm\nu_\mu)\,;~~i=2,3,4\,, 
\end{eqnarray}
so that $S^{\rm eff}_1=\sum\limits_{i=2}^{4} S^{i,\,{\rm eff}}_1$, and
\begin{eqnarray}
  S^{\rm eff}_2\,&:&\,\,\, g^2_{W \neut{1} \,\chgo{2}}\,\times\,{\rm BR}(\chgo{2} \to \chgo{1} \, \mu^+\mu^-)\,\times\, {\rm BR}(\chgo{1} \to \neut{1}\,\mu^\pm\nu_\mu)\,.
  \label{eq:s1eff}
\end{eqnarray}
corresponding to the two signal processes of our interest. Note that the $g^2_{W \neut{i} \, \chgo{j}}$  in the above equations have been simply defined as the sums of the absolute-squares of the left- and right-handed $W^\pm$-neutralino-chargino couplings,
\begin{eqnarray}
  g^2_{W \neut{i} \, \chgo{j}} &\equiv& \left|-ig_2\big(U_{j1}^* N_{i2} + \frac{1}{\sqrt{2}}U_{j2}^* N_{i3}\big)\left(\gamma_\mu \cdot \frac{1-\gamma_5}{2}\right)\right|^2 \nonumber \\
&&+ \left|ig_2\big(\frac{1}{\sqrt{2}}V_{j2} N_{i4}^* - V_{j1} N_{i2}^*\big)\left(\gamma_\mu \cdot \frac{1+\gamma_5}{2}\right)\right|^2\,.
\label{eq:NCWcpl}
\end{eqnarray}
The purpose of defining the dimensionless \snet\ and \scet\ is a qualitative and comparative overview of the two processes without having to calculate their total cross sections for each allowed point. Thus, the neutralino BRs used in eq.~(\ref{eq:s1eff}) actually imply
\begin{eqnarray}
  {\rm BR}(\neut{i} \to \neut{1} \, \mu^+\mu^-)&=&  \left\{ \begin{array}{lr}
    {\rm BR}(\neut{i}\to \neut{1}\,\mu^+\mu^-)\,, & \Delta m_{\neut{i}} < m_Z \\
    {\rm BR}(\neut{i}\to \neut{1}\, Z)\times{\rm BR} (Z\to \mu^+\mu^-)\,, & m_Z < \Delta m_{\neut{i}} < m_h \\
    \sum\limits_{X=Z,\, h}{\rm BR}(\neut{i} \to \neut{1}\,X)\times{\rm BR} (X\to \mu^+\mu^-)
    \,, & m_h < \Delta m_{\neut{i}} \end{array}\right.\,.\nonumber\\
\end{eqnarray}

\begin{figure}[h!]
    \vspace{-1.0cm}
  \begin{tabular}{cc}
    \hspace*{-1.5cm}
    \subfloat[]{
      \includegraphics*[width=9cm]{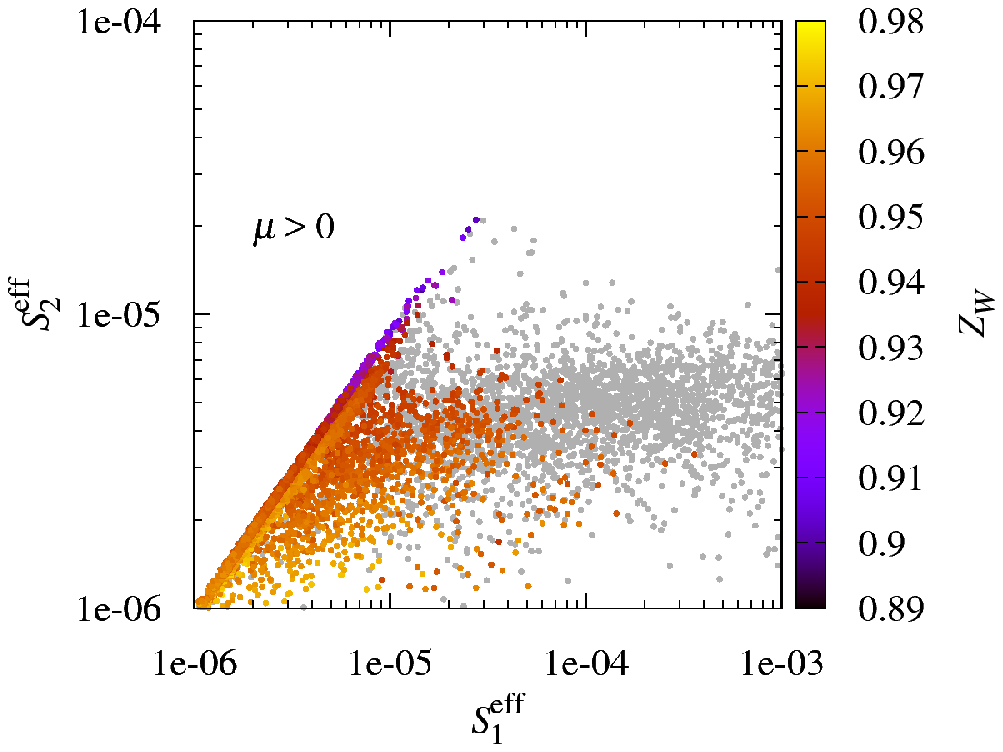}
    } &
    \hspace*{-1.6cm}
    \subfloat[]{
      \includegraphics*[width=9cm]{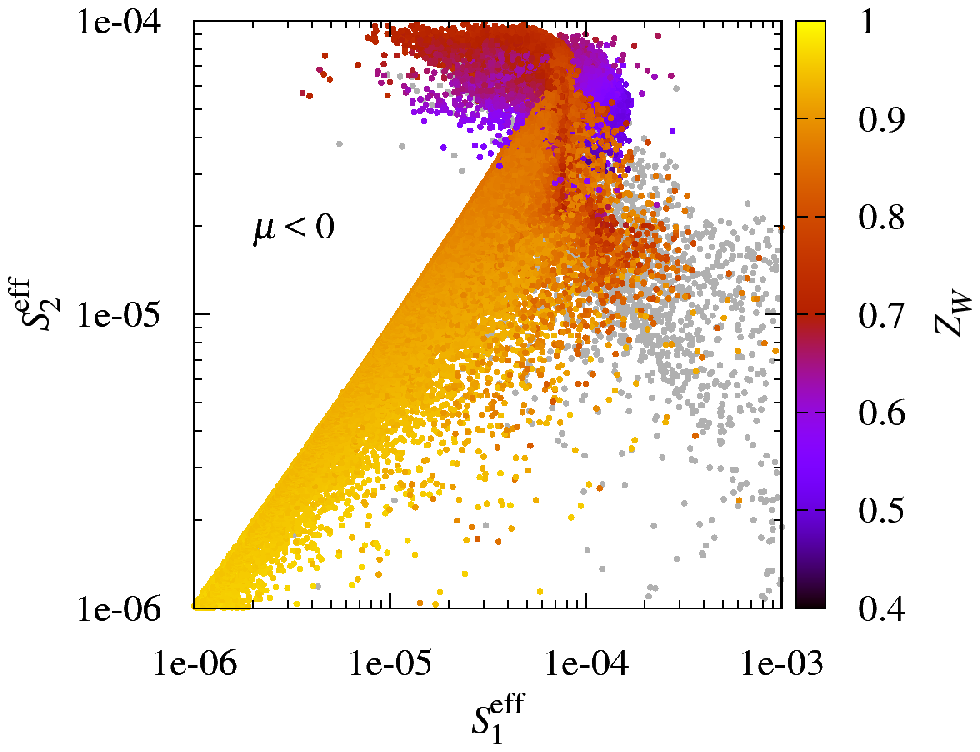}
      } \\
    \hspace*{-1.5cm}
    \subfloat[]{
      \includegraphics*[width=9cm]{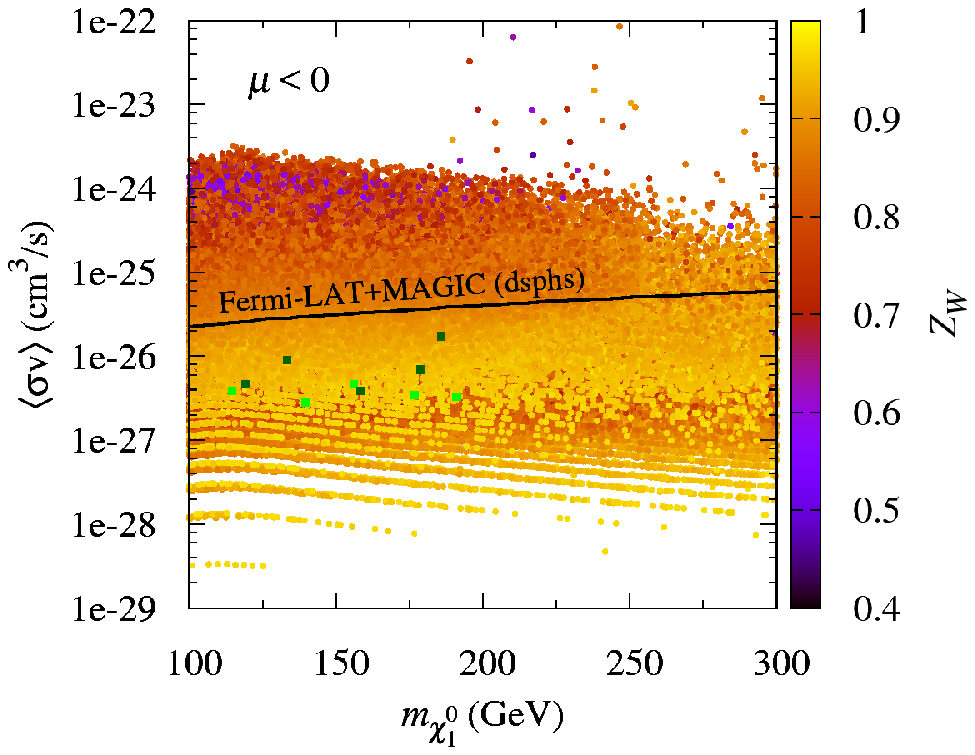}
    } &
    \hspace*{-1.6cm}
    \subfloat[]{
      \includegraphics*[width=9cm]{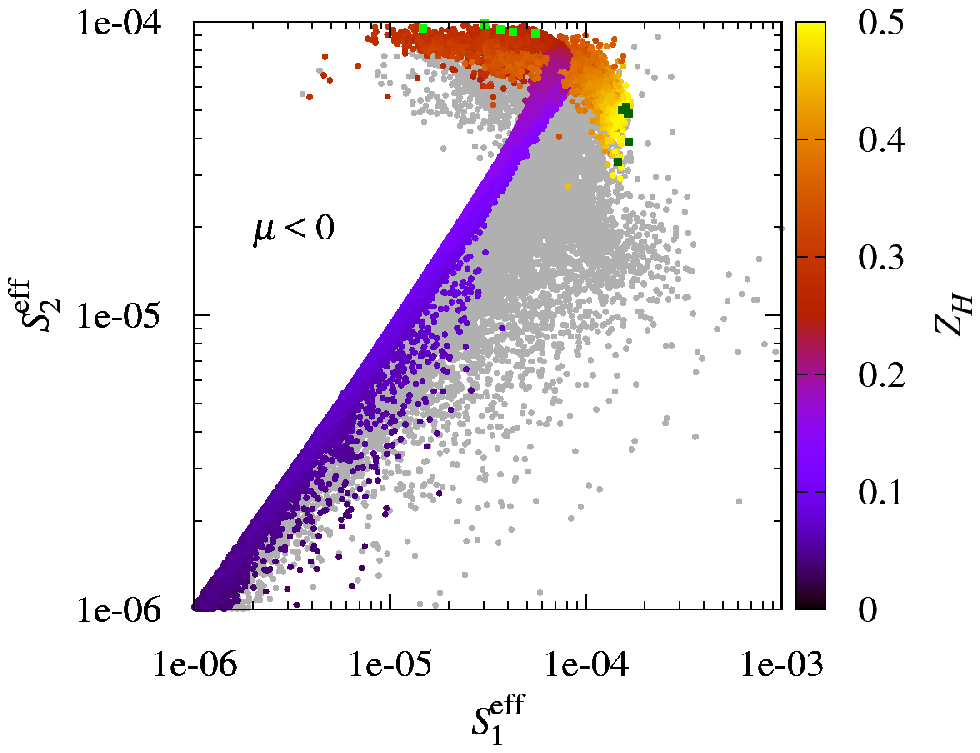}
      }
\end{tabular}
\caption{\label{fig:S1S2} The two effective cross sections, \snet\ ($x$-axis) and \scet\ ($y$-axis), for the allowed points from the parameter space scans for (a) the $\mu >0$ case and (b,\,c,\,d) the $\mu <0$ case. The colour map in (a,\,b and c) corresponds to the wino component and in (d) to the higgsino component in \neut{1}. The grey points in (a,\,b) are the ones ruled out by the XENON1T constraint, and are shown only for a perspective. The grey points in panel (d) are similarly the ones further ruled out by the combined 95\% CL exclusion limits from Fermi-LAT and MAGIC collaborations, i.e, those lying above the corresponding line shown in panel (c).} 
\end{figure}

The top panels of figure~\ref{fig:S1S2} show \snet\ and \scet\ for the scanned points along the x and y axes, respectively. We see in panel (a) that, in the case of $\mu > 0$, points with the largest values of \snet\ (shown in grey in the background) are ruled out by the 95\% confidence level (CL) exclusion limits on $\sigma^{\rm SI}_p$ from the XENON1T direct detection (DD) experiment~\cite{Aprile:2017iyp}, so that for the allowed points \snet\ hardly exceeds $10^{-4}$. \scet\ is comparatively much smaller for this case, only just crossing $10^{-5}$, irrespective of the XENON1T constraint. In contrast, for $\mu < 0$ (shown in panel (b)), some points with $\snet\sim 10^{-3}$ survive the XENON1T limits, while \scet\ too can reach as high as $10^{-4}$ for a considerable proportion of the allowed points. Therefore, we will limit our analysis from here onwards only to points belonging to the $\mu < 0$ case.

In panel (c) we show the consistency of the remaining points against currently the strongest 95\% CL exclusion limits from indirect detection (ID) for DM masses of our interest, which come from the combined analysis of $\neut{1}\neut{1} \to b\bar{b}$ annihilation in dwarf spheroidel galaxies (dSphs) performed by the Fermi-LAT gamma-ray telescope and the MAGIC Cherenkov detector collaborations~\cite{Ahnen:2016qkx}. One sees that this constraint further rules out a large portion of points that satisfy the DD limits. According to panel (d), the only points for which \snet\ approaches about $10^{-3}$ are the excluded (grey) ones, for which $M_2 \lesssim M_1$ while $|\mu|$ is relatively large, implying that there is a significant bino component in \neut{1}\ and the \neut{2} is bino-like.

Among the points surviving both the above constraints,
when the DM is almost purely wino, which can be identified from the colour maps showing $Z_W$ and $Z_H$ in panels (b) and (d), respectively, of figure~\ref{fig:S1S2}, the two effective cross sections typically stay below $10^{-5}$. 
For points with wino-higgsino DM and higgsino-like \neut{2,3}, lying in the top 1/4th of the area of the two panels, \scet\ is highly enhanced. These points can in fact be divided into two scenarios. In the first one, \neut{1}\ is almost equal parts wino and higgsino (the purple points in panel (b)). In this scenario \sne{2}\ and \scet\ mostly have nearly identical values, and we refer to it as the `wino with large higgsino' (WLH) scenario here. In the second, `wino with small higgsino' (WSH), scenario, $Z_W$ in \neut{1}\ dominates over $Z_H$ by roughly 7:3 (the red points in panel (b)). Crucially for this scenario, the \scet\ can be more than an order of magnitude larger than \snet. The points highlighted in light (dark) green in panels (c) and (d) are the BPs corresponding to the WSH (WLH) scenario that we selected for our detector-level analysis discussed later. 

\begin{figure}[t!]
    \vspace{-.5cm}
  \begin{tabular}{cc}
\hspace*{-1.5cm}
    \subfloat[]{
      \includegraphics*[width=9cm]{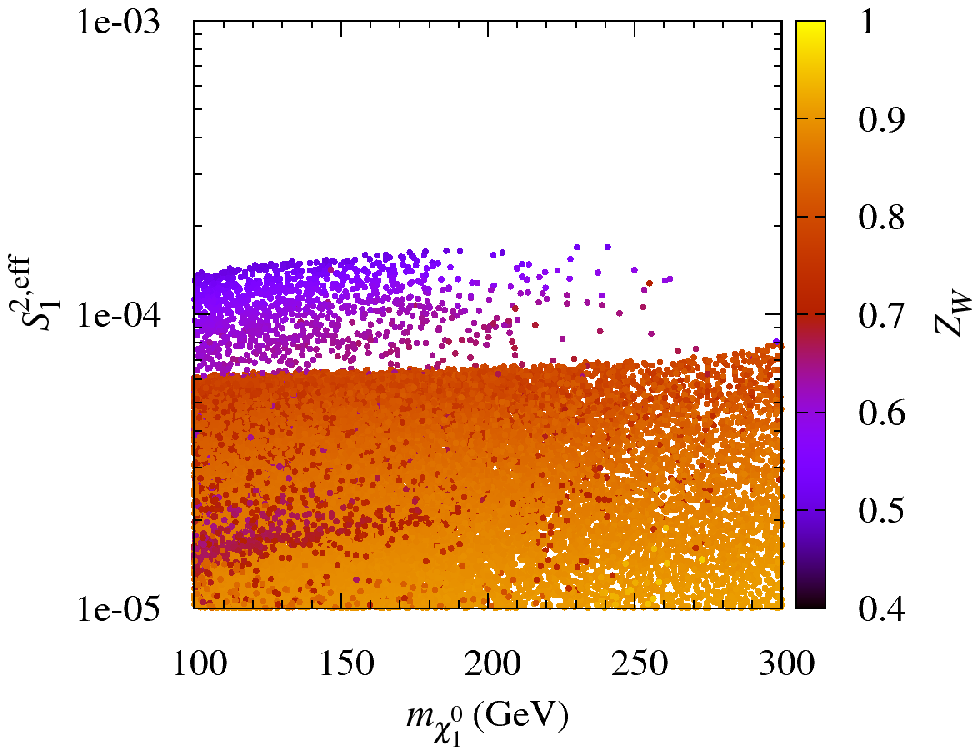}
          } &
    \hspace*{-1.8cm}
    \subfloat[]{
      \includegraphics*[width=9cm]{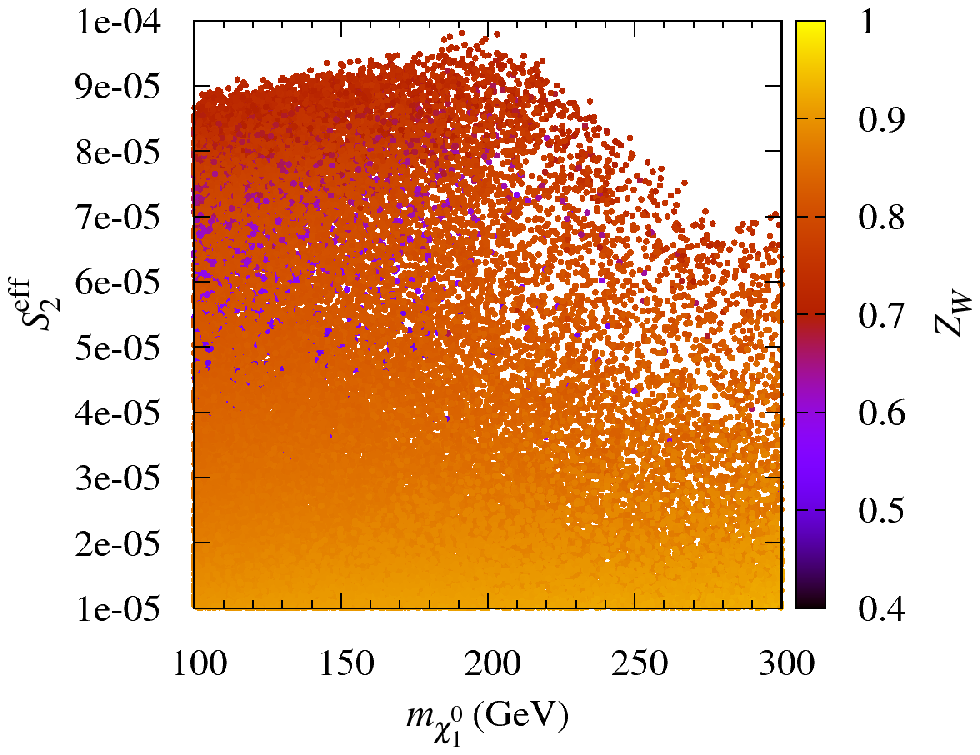}
      }\\
    \hspace*{-1.5cm}
    \subfloat[]{
      \includegraphics*[width=9cm]{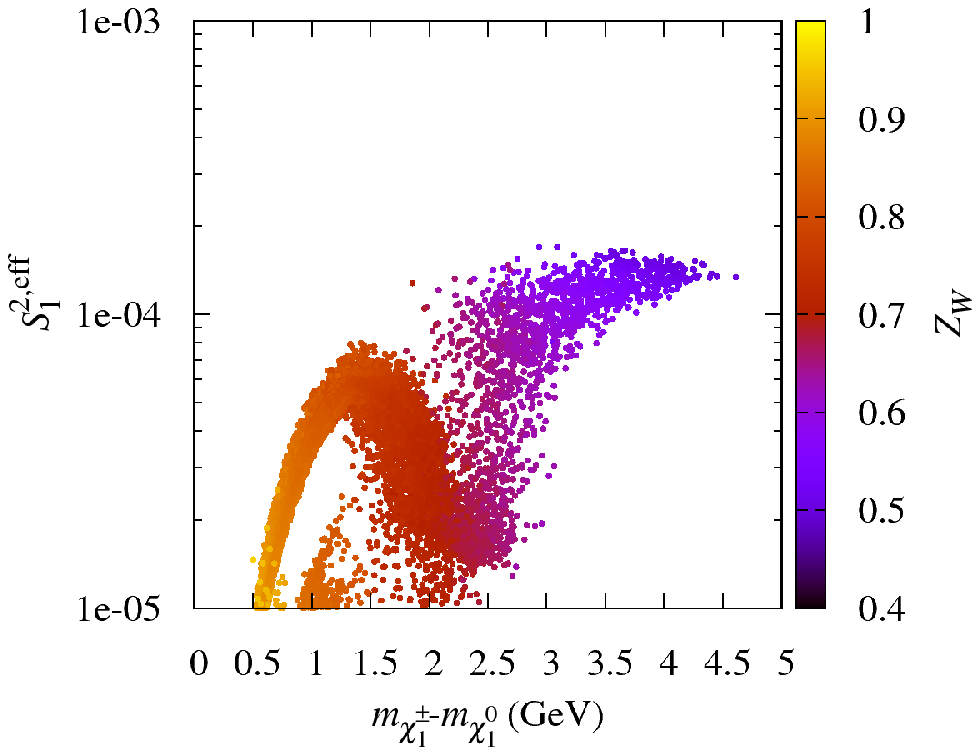}
          } &
    \hspace*{-1.8cm}
    \subfloat[]{
      \includegraphics*[width=9cm]{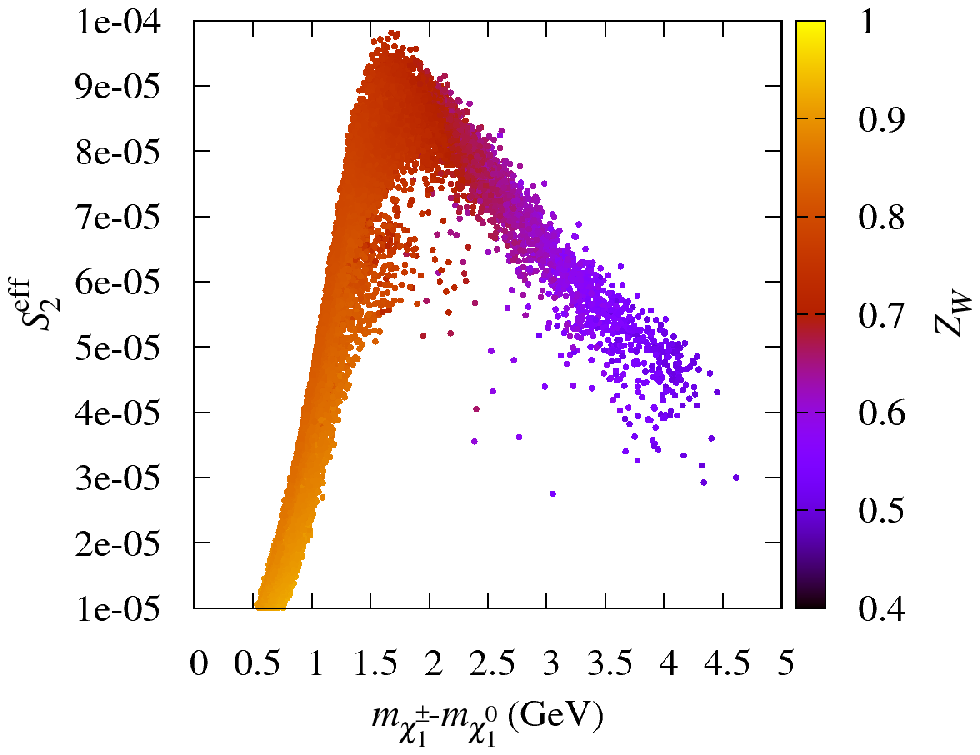}
      }
    \end{tabular}
\caption{\label{fig:m-delmcn} (a) \sne{2} and (b) \scet\ as functions of the \neut{1} mass. In panels (c) and (d) the same quantities are plotted, respectively, as functions of the \neut{1}-\chgo{1} mass difference. The colour maps in all the panels correspond to the wino component in \neut{1}.} 
\end{figure}

In figure~\ref{fig:m-delmcn} (and onwards), we show only \sne{2}\ for further clarity, as it is the dominant contributor to \snet. We notice in panel (a) that the maximal \sne{2}\ stays almost constant with increasing $m_{\neut{1}}$, while according to panel (b) \scet\ rises slowly until $m_{\neut{1}}$ reaches about 200\,GeV and then falls relatively steeply. The bottom panels of figure~\ref{fig:m-delmcn} show that the maximal value of \sne{2}\ (\scet) is obtained for maximal (minimal) $\mchgo{1} - \mneut{1}$ possible in each scenario (given the experimental constraints imposed). However, a conflict with the strong CMS limit on the chargino lifetime~\cite{CMS:2014gxa} does not arise, since this $\mchgo{1} - \mneut{1}$ is always higher than 1\,GeV~\cite{Han:2015lma}, except for some points with a very large $Z_W$ (which we will ignore here). 

\begin{figure}[t!]
    \vspace{-0.5cm}
  \begin{tabular}{cc}
\hspace*{-1.5cm}
    \subfloat[]{
      \includegraphics*[width=9cm]{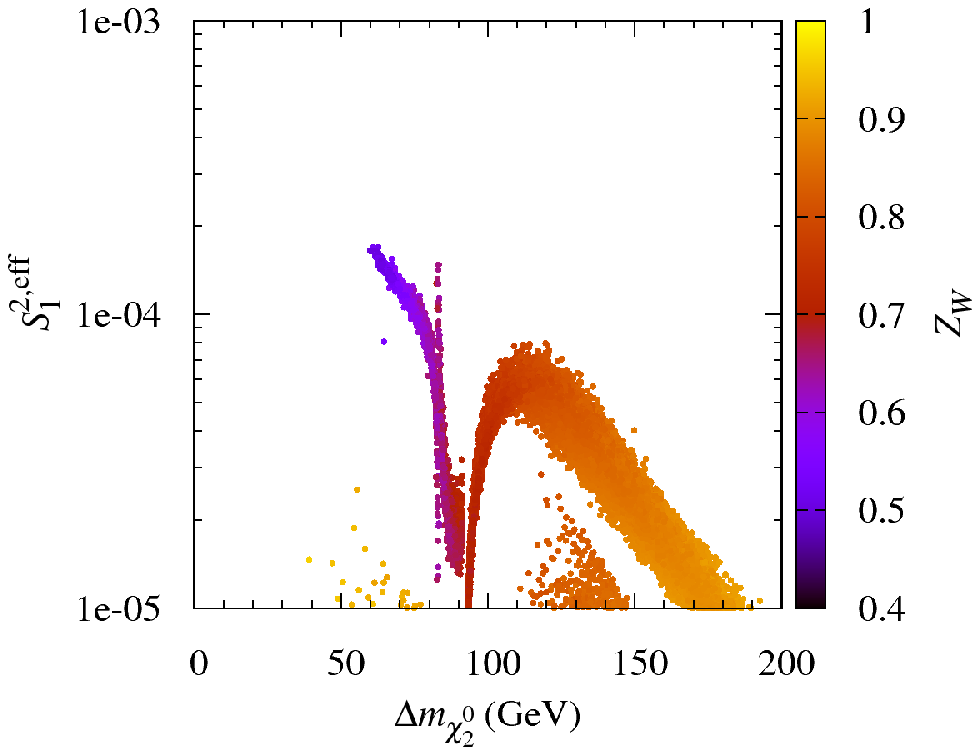}
                } &
    \hspace*{-1.8cm}
    \subfloat[]{
      \includegraphics*[width=9cm]{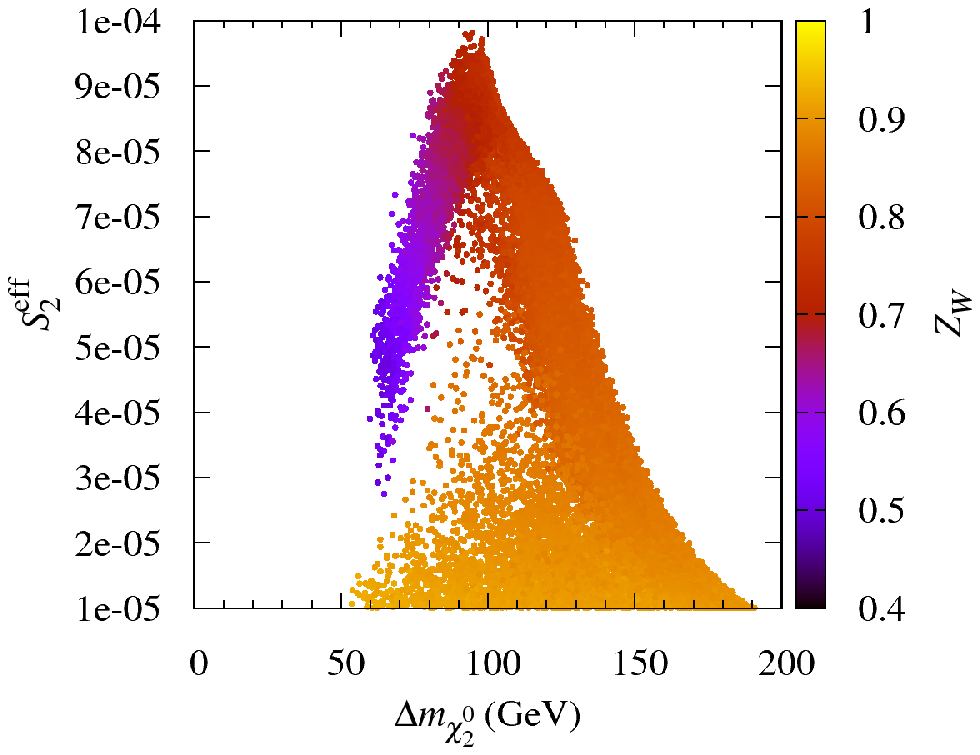}
    } \\
    \hspace*{-1.5cm}
    \subfloat[]{
      \includegraphics*[width=9cm]{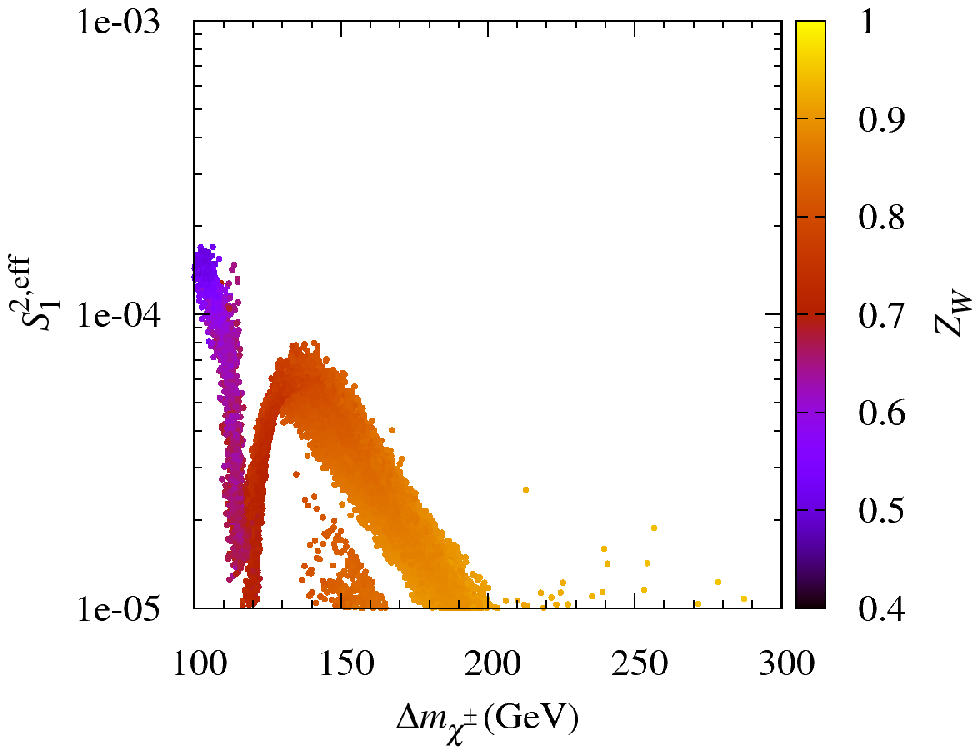}
                } &
    \hspace*{-1.8cm}
    \subfloat[]{
      \includegraphics*[width=9cm]{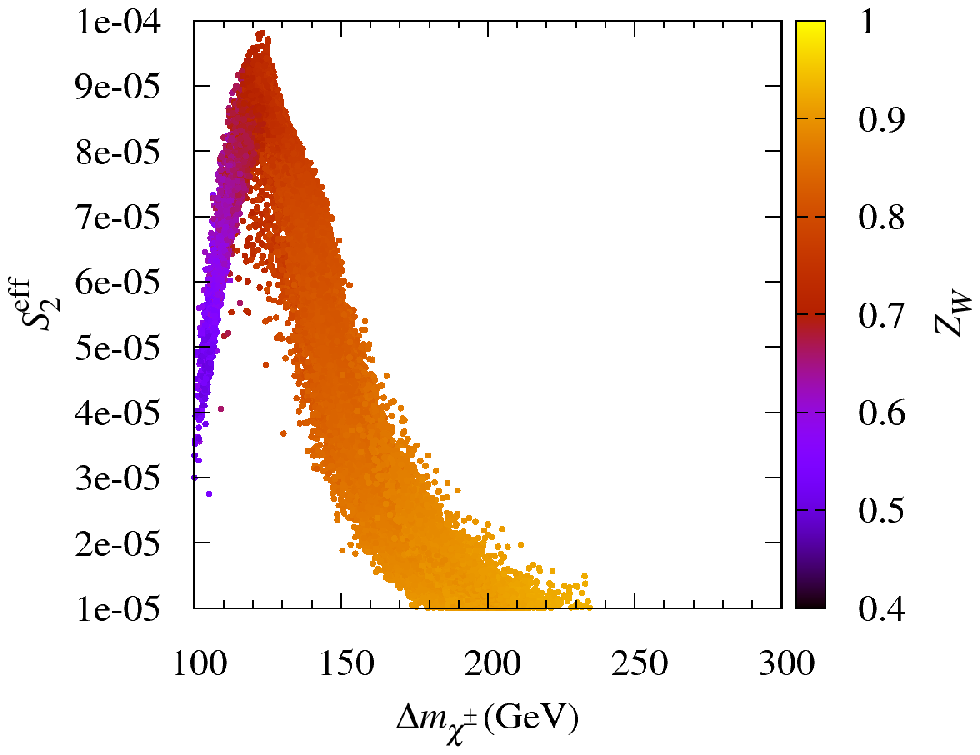}
      }
    \end{tabular}
\caption{\label{fig:delmn-delmc} (a) \sne{2} and (b) \scet\ as functions of the \neut{1}-\neut{2} mass difference. In panels (c) and (d) the same quantities are plotted, respectively, as functions of the \chgo{1}-\chgo{2} mass difference. The colour maps in all the panels correspond to the wino component in \neut{1}.} 
\end{figure}

The most important feature of relevance to this study is illustrated in figure~\ref{fig:delmn-delmc}, where we note that points with $\dmn{2} \lesssim 60$\,GeV do not appear. This is because a larger higgsino component in \neut{1}\ leads to exclusion by the XENON1T constraint, implying that \mneut{1}\ lying just above the noted value only barely satisfy it. With increasing $Z_W$ the mass-splitting increases and \sne{2} first drops slowly and then takes a sharp dip around the $\neut{2}\to \neut{1} Z$ threshold. It is precisely at this \dmn{2}\ that the transition from the WLH to the WSH scenario takes place, so that \scet\ becomes dominant over \sne{2}, as noted earlier and seen again in figure~\ref{fig:delmn-delmc}(b). In fact, \scet\ has the maximal achievable values here, and thus compensates for the reduction in \sne{2}. According to figure~\ref{fig:delmn-delmc}(c), when \dmn{2}\ is around the $\neut{1} Z$ threshold (causing the dip in \sne{2}), $\dmc$ lies above $\sim 100$\,GeV, so that the $\chgo{2} \to \chgo{1} Z$ decay is allowed. This results in the enhancement in \scet\ seen in panel (d), until \dmc\ exceeds $\sim 120$\,GeV, after which the $\chgo{2} \to \chgo{1} h$ decay channel opens up, pulling \scet\ down.

\begin{figure}[t!]
    \vspace{-0.5cm}
  \begin{tabular}{cc}
\hspace*{-1.5cm}
    \subfloat[]{
      \includegraphics*[width=9cm]{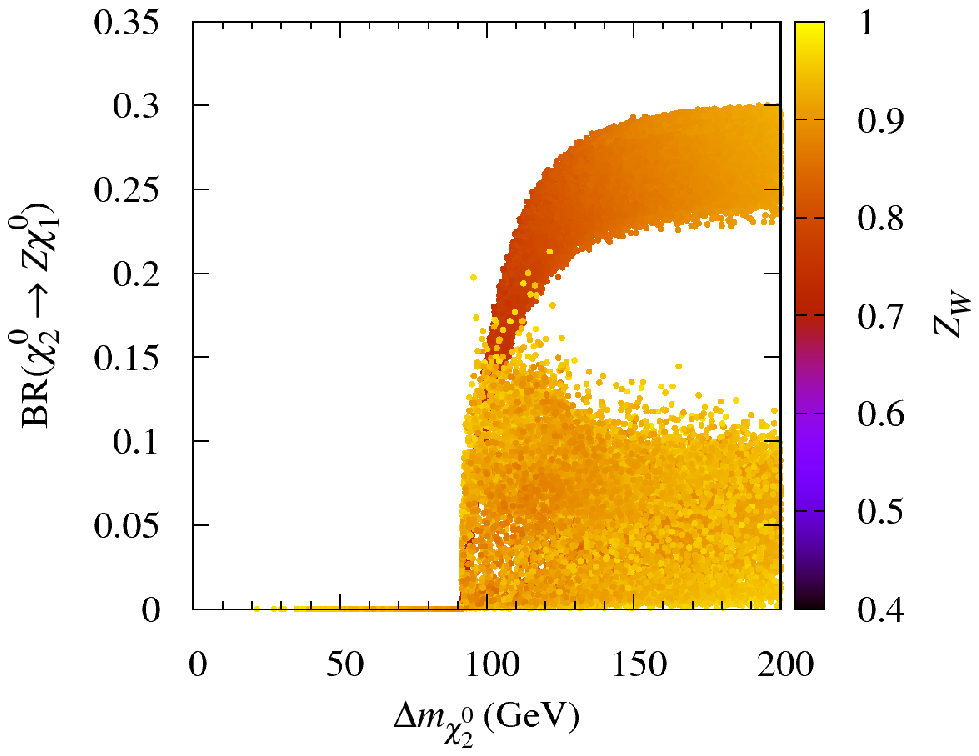}
                      } &
    \hspace*{-1.8cm}
    \subfloat[]{
      \includegraphics*[width=9cm]{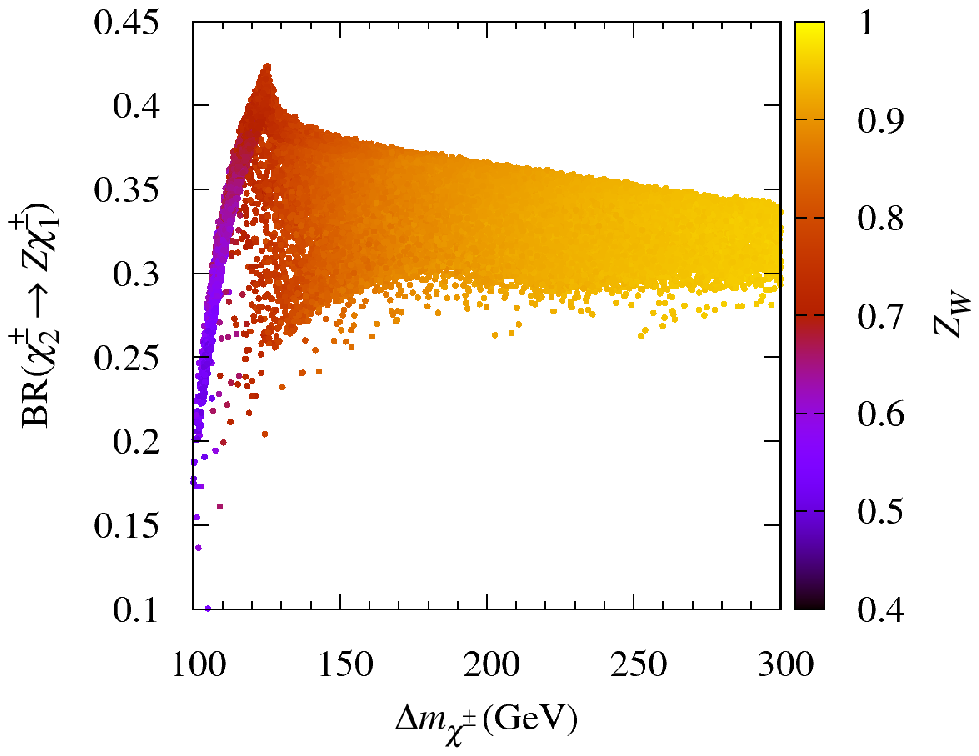}
      }
    \end{tabular}
\caption{\label{fig:BRs} (a) BR($\neut{2}\to \neut{1} Z$) as a function of the \neut{1}-\neut{2} mass difference, and (b) BR($\chgo{2}\to \chgo{1} Z$) as a function of the \chgo{1}-\chgo{2} mass difference. The colour map corresponds to the wino component in \neut{1}.} 
\end{figure}

The behaviour of \sne{2} and \scet\ discussed above is evidently driven largely by the BRs for the $\neut{2} \to \neut{1} Z$ and $\chgo{2} \to \chgo{1} Z$ decays, which are shown in figure~\ref{fig:BRs}, panels (a) and (b), respectively. BR($\neut{2}\to \neut{1} Z$) is minimal right at the threshold, which is a consequence of the corresponding coupling,
\begin{eqnarray}
  g^2_{Z\neut{i} \, \neut{j}} &=& \left|-\frac{i}{2}\big(g_1\sin\theta_W + g_2 \cos\theta_W\big)\big(N_{j3}^* N_{i3} - N_{j4}^* N_{i4}\big)\left(\gamma_\mu \cdot \frac{1-\gamma_5}{2}\right)\right|^2  \nonumber \\
&&+ \left|\frac{i}{2}\big(g_1\sin\theta_W + g_2 \cos\theta_W\big)\big(N_{i3}^* N_{j3} - N_{i4}^* N_{j4}\big)\left(\gamma_\mu \cdot \frac{1+\gamma_5}{2}\right)\right|^2\,,
\label{eq:NNZcpl}
\end{eqnarray}
depending only on the higgsino components in \neut{1}\ and \neut{2}. Thus, for a wino-like \neut{1}\ and higgsino-like \neut{2,3} this coupling can undergo reduction due to large cancellations. Such cancellations are not as strong in the case of the coupling,
\begin{eqnarray}
  g^2_{Z\chgo{i} \, \chgo{j}} &=& \left|\frac{i}{2}\bigg(2g_2U_{j1}^*\cos\theta_W U_{i1} + U_{j2}^*\big(-g_1\sin\theta_W + g_2 \cos\theta_W\big)U_{i2}\bigg)\left(\gamma_\mu \cdot \frac{1-\gamma_5}{2}\right)\right|^2 \nonumber \\
&& + \left|\frac{i}{2}\bigg(2g_2V_{i1}^*\cos\theta_W V_{j1} + V_{i2}^* \big(-g_1\sin\theta_W + g_2 \cos\theta_W\big)V_{j2}\bigg)\left(\gamma_\mu \cdot \frac{1+\gamma_5}{2}\right)\right|^2\,,\nonumber \\
\label{eq:CCZcpl}
\end{eqnarray}
responsible for the $\chgo{2}\to \chgo{1} Z$ decay, the BR for which consequently rises to about 42\% just before the $\chgo{2} \to \chgo{1} h$ decay threshold. We point out here that the $\neut{3}\to \neut{1} Z$ decay follows a characteristic trend very similar to the $\neut{2}\to \neut{1} Z$ decay in this scenario, and is therefore not illustrated separately here.

\section{\label{sec:analysis} Signal-to-background analysis of benchmark points}
In order to perform a detector-level analysis, we picked five BPs for each of the two DM scenarios from the allowed points, as noted earlier. In the WSH (WLH) scenario, those points were identified as BPs for which the highest \scet\ (\snet) was obtained in each of the five bins of 20\,GeV in \mneut{1} ranging between 100--200\,GeV. This upper cut-off at 200\,GeV in \mneut{1} is essentially inspired, besides the phase-space considerations, by the observation in figure~\ref{fig:m-delmcn}(b) that the \scet\ starts declining rapidly after this value. For the selected BPs, the parton-level cross sections at the leading order (LO) for the $S_1$ and $S_2$ signals were calculated with {\tt MadGraph5\_aMC@NLO}~\cite{Alwall:2014hca} for the LHC with $\sqrt{s}=14$\,TeV.\footnote{We have adopted here 14\,TeV as a realistically achievable energy by the LHC over the time frame necessary to accrue the ${\cal O}(1000)$ fb$^{-1}$ luminosities needed to establish some of our signals (as we will see below). However, if we were to instead use $\sqrt{s}=13$\,TeV in our analysis, the cross sections would typically be lower by $10-20\%$ and the signal significances would scale accordingly.} These cross sections were then multiplied by a $k$-factor of 1.25, since it is almost a constant for the neutralino and chargino mass ranges of our interest, obtained from the public code {\tt PROSPINO-v2.1}~\cite{Beenakker:1996ed}. As noted in the previous section, for a given \mneut{1}\ in each scenario, \sne{2}\ and \scet\ peak for particular values of \dmn{2}\ and \dmc. Thus, once a scenario and \mneut{1}\ in it have been specified, \mneut{2}\ and \mchgo{2}\ are almost fixed by the requirement of maximal \sne{2}\ and \scet. For this reason, in table~\ref{tab:BPs} we provide the neutralino and chargino masses along with the $S_1$ and $S_2$ cross sections only for the lowest \mneut{1}\ BPs chosen for each of the two scenario, for reference in the discussion to follow. 

\begin{table}[tbp]
\begin{center}
 \begin{tabular}{|c|c|c|c|c|c|c|c|c|}
    \hline            
\multirow{2}{*}{BP (scenario)} & 
$m_{{\tilde{\chi}}^0_1}$&$m_{{\tilde{\chi}}^0_2}$&$m_{{\tilde{\chi}}^0_3}$&$m_{{\tilde{\chi}}^0_4}$&$m_{{\tilde{\chi}}^\pm_1}$&$m_{{\tilde{\chi}}^\pm_2}$&$\sigma_{S_1}$&$\sigma_{S_2}$
\\ 
&
[{\rm GeV}]&[{\rm GeV}]&[{\rm GeV}]&[{\rm GeV}]&[{\rm GeV}]&
[{\rm GeV}]&[fb]& [fb] \\
\hline \hline
1 (WSH)&115 & 214&234 & 891& 117&243 &17 & 18.7 \\
\hline
2 (WLH)&119 & 183&216 & 844& 123&223 & 63.7 & 9.66\\
\hline
\end{tabular}
\end{center}
\caption{\label{tab:BPs} Masses of the neutralinos and charginos as well as cross sections corresponding to the two signal processes for the BP with the lowest \mneut{1}\ in each scenario.}
\end{table}

The SM backgrounds ($B$) were also obtained from {\tt MadGraph}, but at the next-to-LO (NLO) for consistency. The {\tt MadGraph} outputs were then passed to {\tt PYTHIA6}~\cite{Sjostrand:2006za} for parton-showering and hadronisation, with jet ({\it j})-clustering (where $j=g,u,c,d,s,b$) performed using the anti-$k_t$ algorithm~\cite{Cacciari:2008gp} with a cone radius $\Delta R = 0.5$, pseudorapidity $|\eta(j)| < 2.5$, and $p_T(j) > 30$\,GeV, and subsequently to the {\tt DELPHES} package~\cite{deFavereau:2013fsa} for fast detector simulation. Finally, we manipulated the Monte Carlo (MC) data with {\tt MadAnalysis5}~\cite{Conte:2012fm}. 

\begin{figure}[t!]
\begin{center}
\includegraphics*[width=7.5cm]{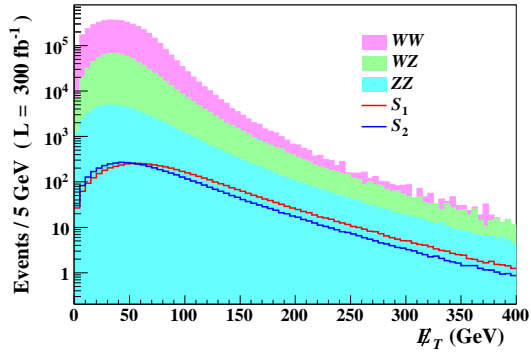}
\end{center}
\caption{\label{fig:MET} Cumulative number of signal and background events for BP1 versus $\met$ at the LHC with $\sqrt{s}=14$\,TeV for $\mathcal{L}=300$\,fb$^{-1}$.} 
\end{figure}

We required each event to contain exactly three reconstructed leptons, with each of them having $|\eta(\ell)|<2.4$, two being OSSF ones, and at least one having $p_T>20$\,GeV. Jets were separated from the lepton candidates through $\Delta R(\ell,j)>0.4$, and those consistent with anomalous noise in the calorimeters were rejected~\cite{Yetkin:2011zz}. The dominant irreducible backgrounds for this search are from $W^\pm Z$ and $t\bar{t}$ production, of which the latter is suppressed by rejecting events with at least one $b$-tag. We also take into account the subdominant background from $ZZ$ production, but disregard those from rare SM processes such as $t\bar{t}Z$, $t\bar{t}W$, $t\bar{t}H$ and tri-boson production. Furthermore, since the DM of our interest here is wino-like and thus always heavier than $\sim 100$\,GeV, rejecting events with $\met < 100$\,GeV can substantially suppress the additional backgrounds from events with $Z+{\rm jets}$ and  $W^+W^-$ production. This can be easily deduced from figure~\ref{fig:MET}, containing the $\met$ distributions for the two signals and the considered backgrounds. 

\begin{figure}[h!]
\begin{center}
    \subfloat[]{
\includegraphics*[width=7.5cm]{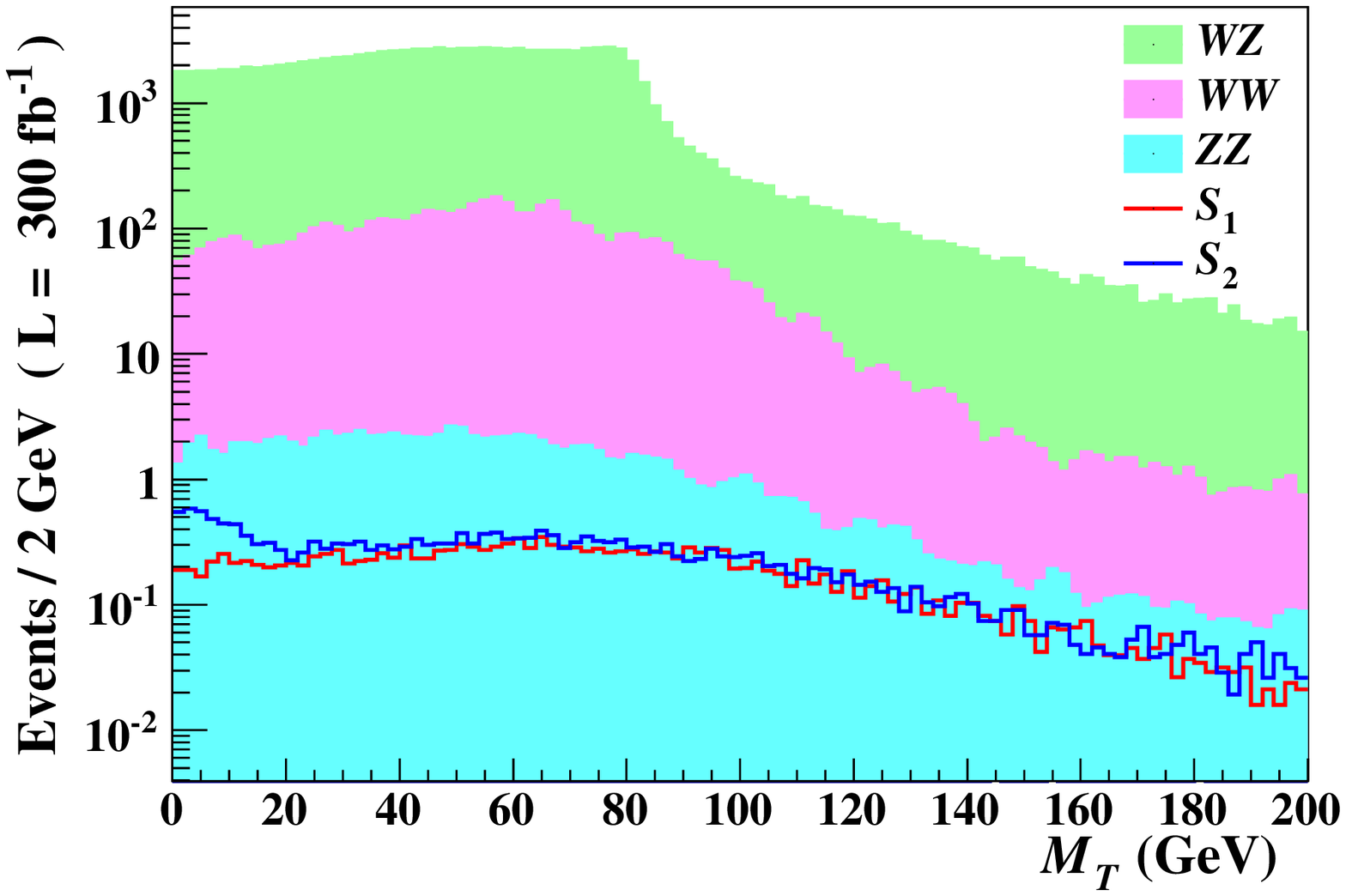}
}
    \subfloat[]{
\includegraphics*[width=7.5cm]{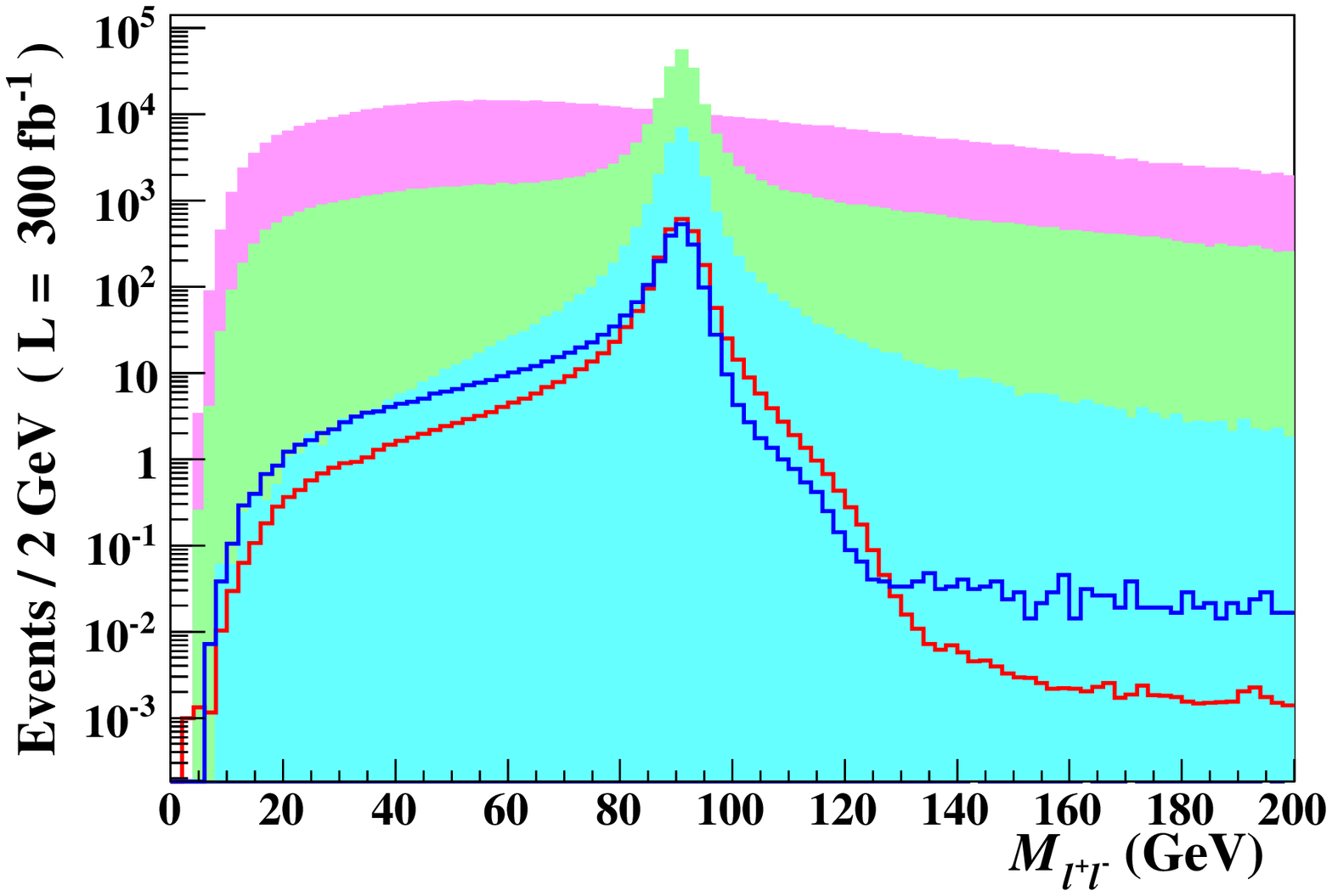}
}
\end{center}
\caption{\label{fig:MllMT} Cumulative number of signal and background events for BP1 versus (a) the invariant mass of the OSSF lepton pair and (b) the transverse mass at the LHC with $\sqrt{s}=14$\,TeV for $\mathcal{L}=300$\,fb$^{-1}$. The legend in (a) applies to (b) also.} 
\end{figure}

In the trilepton searches at the LHC, the two OSSF leptons are characterised by the invariant mass $M_{\ell^+ \ell^-}$, while the third lepton, $\ell_3$, is identified by constructing the transverse mass,
\begin{equation}
  M_T =\sqrt{2\,\met\,p_{T(\ell_3)}\,(1-\cos \Delta \phi_{\ell_3,\met})}\,,
\end{equation}
where $\Delta \phi_{\ell_3,\met}$ is the difference between the azimuthal angles of $\ell_3$ and $\met$. The LHC searches~\cite{CMS:2017sqn,CMS:2017fdz,Khachatryan:2014qwa,ATLAS:2017uun,ATLAS:2016uwq,Aad:2014nua} define certain regions in $M_T$, i.e., $M_T$ $>160$\,GeV, $120$\,GeV $<M_T<$ $160$\,GeV, and $0$\,GeV  $<M_T<$ $120$\,GeV, in order to suppress the huge $W^\pm Z$ background. In the wino-like DM scenario though, the near mass-degeneracy between \neut{1}\ and \chgo{1}\ implies that the $\ell_3$ is generally very soft, so that any cut on $M_T$ almost completely diminishes both the signals, as figure~\ref{fig:MllMT}(a) illustrates. However, figure~\ref{fig:MllMT}(b) shows that a $75~\text{GeV}<M_{\ell^+ \ell^-}<105~\text{GeV}$ selection condition on the OSSF lepton pair can prove crucial in recovering the signal events while keeping the SM background under control.

\begin{table}[tbp]
    \resizebox{\textwidth}{!}{
      \begin{tabular}{|c||c|c|c|c|c|c|c|c|c|}
      \hline            
            \multicolumn{2}{|c|}{ \multirow{2}{*}{Cuts}}& \multicolumn{3}{c|}{Backgrounds}& \multicolumn{2}{c|}{Signals}&\multicolumn{3}{c|}{Significances}\\    
      \cline{3-10}       
\multicolumn{2}{|c|}{}&  $W^\pm Z$ & $W^+W^-$ & $ZZ$ & $S_1$ & $S_2$ & $\mathcal{Z}_1$ & $\mathcal{Z}_2$&  $\mathcal{Z}_{3\ell}$\\
     \hline  \hline                           
\multicolumn{2}{|c|}{Events before cuts} &              778670 & 4444650 & 73213 & 5084 & 5616 & 2.21 & 2.44 &4.65 \\ 
     \hline
\multicolumn{2}{|c|}{$|\eta(\ell)|<2.4$}&                  701373 & 3538624 & 59515 & 4537 & 5046 & 2.19 & 2.43 &4.62 \\ 
     \hline
\multicolumn{2}{|c|}{$n(\ell)\geq 1$ with $p_T > 20$\,GeV}& 674906 & 3295789 & 56633 & 4366 & 4902 & 2.18 & 2.44 &4.62 \\
      \hline
\multicolumn{2}{|c|}{$\Delta R(\ell,j)>0.4$}&              428099 & 1616502 & 27309 & 2543 & 2887 & 1.77 & 2.01 &3.78 \\
      \hline
\multicolumn{2}{|c|}{$p_T(j) > 30$\,GeV}&                343438 &  1280660 & 21741 & 2154 & 2457 & 1.68 & 1.92 &3.60 \\
      \hline
\multicolumn{2}{|c|}{$|\eta(j)|<2.5$}&                  270703 &  1017195 & 18403 & 1943 & 2222 & 1.70 & 1.94 &3.64 \\
     \hline
\multicolumn{2}{|c|}{$b$-jet veto}&                    267997 &  984850 & 17870 & 1859 & 2127 & 1.65 & 1.89 &3.54 \\                
\hline
\hline
\rowcolor{red!25}
\multirow{3}{*}{}&$>160$\,GeV&                 516 &      16 &     2 &    1 &    1 & 0.04 & 0.04& 0.08 \\                
      \cline{2-10} 
\rowcolor{red!25}           
$M_T$  &$120-160$\,GeV&               750 &      54 &     2 &    1 &    1 & 0.04 & 0.04& 0.08 \\                
      \cline{2-10}
\rowcolor{red!25}
                                 &$0-120$\,GeV&               35705 &    1827 &    34 &    9 &    7 & 0.05 & 0.04& 0.09 \\                
      \hline
      \hline
      \rowcolor{green!25}    
\multicolumn{2}{|c|}{\cellcolor{green!25}$\met >100$\,GeV}&21026 &   60391 &  3411 &  666 & 917 & 2.29 & 3.15 &5.44 \\
      \hline
      \rowcolor{green!25}    
\multicolumn{2}{|c|}{\cellcolor{green!25}  $75\,\text{GeV}<M_{\ell^+ \ell^-}<105\,\text{GeV}$}
                                                        & 7452 &    2793 &  1511 & 319 &  459 & 2.94 & 4.23 & 7.17 \\
\hline
  \end{tabular}
  }
\caption{\label{tab:300fb} Cut-flow for $S_1$, $S_2$ and the SM backgrounds for BP1, at the LHC with $\sqrt{s}=14$\,TeV for $\mathcal{L}=300$\,fb$^{-1}$. The numbers in the last two (green-shaded rows) are obtained by disregarding the cuts on $M_T$ (shown in the red-shaded rows).}   
\end{table}

For each of the ten BPs we then calculated the statistical significance, defined as 
\begin{eqnarray}
\mathcal{Z}_i \equiv \frac{N_{S_i}}{\sqrt{N_B}}\,,
\label{eq:significance}
\end{eqnarray}
where the number of events for the $S_1$ signal is given as $N_{S_1}=\sum\limits_{i=2}^{4} N_{S^i_1}$, with $N_{S^i_1}$ corresponding to the $S^i_1$ signal processes for $i=1-3$, $N_{S_2}$ is the number of $S_2$ signal events  and $N_B$ the total number of background events. We also define the total number of $3\ell + \met$ signal events as $N_{S_{3\ell}}=N_{S_1}+N_{S_2}$ in order to calculate the combined significance of the two production modes considered.

In table~\ref{tab:300fb}, we show the cut-flow for $N_{S_1}$, $N_{S_2}$ and $N_B$  for BP1, as well as the individual and combined signal significances at the LHC with $\sqrt{s}=14$\,TeV for $\mathcal{L}=300$\;fb$^{-1}$. In accordance with the discussion above, for the final $\mathcal{Z}_i$ quoted, we ignore the $M_T$ selection cuts shown in the red rows, as they highly suppress the signals, and instead impose the following two cuts: $\met>100$\,GeV and $75\,\text{GeV}<M_{\ell^+\ell^-}<105\,\text{GeV}$, in the rows highlighted in green. In table~\ref{tab:1000fb} we show a similar cut-flow for the same BP but with $\mathcal{L}=1000$\,fb$^{-1}$. We note in both the tables that $\mathcal{Z}_2$ is larger than $\mathcal{Z}_1$ and hence our slight modification of the selection procedure raises the combined significance, $\mathcal{Z}_{3\ell}$, to a considerably higher value than what would be achievable with the cuts adopted in the $3\ell + \met$ searches at the LHC so far. 

\begin{table}[tbp]
    \resizebox{\textwidth}{!}{
      \begin{tabular}{|c||c|c|c|c|c|c|c|c|c|}
      \hline            
            \multicolumn{2}{|c|}{ \multirow{2}{*}{Cuts}}& \multicolumn{3}{c|}{Backgrounds}& \multicolumn{2}{c|}{Signals}&\multicolumn{3}{c|}{Significances}\\    
      \cline{3-10}       
\multicolumn{2}{|c|}{}&  $W^\pm Z$ & $W^+W^-$ & $ZZ$ & $S_1$ & $S_2$ & $\mathcal{Z}_1$ & $\mathcal{Z}_2$&  $\mathcal{Z}_{3\ell}$\\
     \hline  \hline                           
\multicolumn{2}{|c|}{Events before cuts} &              2595569 & 14815500 & 244045 & 16948 & 18720 & 4.03& 4.46 &8.49 \\ 
     \hline
\multicolumn{2}{|c|}{$|\eta(\ell)|<2.4$}&                  2337910 & 11795414 & 198385 & 15123& 16821  & 3.99& 4.44 &8.43 \\ 
     \hline
\multicolumn{2}{|c|}{$n(\ell)\geq 1$ with $p_T > 20$\,GeV}& 2249689 & 10985966 & 188777 & 14555 & 16340  & 3.97& 4.46 &8.43 \\
      \hline
\multicolumn{2}{|c|}{$\Delta R(\ell,j)>0.4$}&              1426997 & 5388342 & 91031 & 8478 & 9625 & 3.23& 3.66 &6.89 \\
      \hline
\multicolumn{2}{|c|}{$p_T(j) > 30$\,GeV}&                1144796 &  4268867 & 72470 & 7180 & 8191 & 3.07 & 3.50 &6.57 \\
      \hline
\multicolumn{2}{|c|}{$|\eta(j)|<2.5$}&                  902344 &  3390651 & 61346 & 6476 & 7407 & 3.10 & 3.55 &6.65 \\
     \hline
\multicolumn{2}{|c|}{$b$-jet veto}&                    893325 &  3282834 & 59569 & 6197 & 7090 & 3.01 & 3.45 &6.46 \\                
     \hline\hline
      \rowcolor{red!25}    
\multirow{3}{*}{}&$>160$\,GeV&                 1720 &      55 &     5 &    3 &    3 & 0.07 & 0.07 &0.14 \\                
      \cline{2-10}
      \rowcolor{red!25}    
 $M_T$ &$120-160$\,GeV&               2499 &      179 &    8 &    4 &    4 & 0.08 & 0.08 &0.16 \\                
      \cline{2-10}
      \rowcolor{red!25}    
                                 &$0-120$\,GeV&               119017 &    6090 &    113 &    29 &  23 & 0.08 & 0.07&0.15 \\                
      \hline \hline
      \rowcolor{green!25}    
\multicolumn{2}{|c|}{\cellcolor{green!25} $\met >100$\,GeV}&70087 &   201303 &  11370 &  2220 &  3057  & 4.17 & 5.75&9.92 \\
      \hline
      \rowcolor{green!25}    
\multicolumn{2}{|c|}{\cellcolor{green!25} $75\,\text{GeV}<M_{\ell^+ \ell^-}<105\,\text{GeV}$}
                                                        & 24837 &    9310 &  5039 & 1061 & 1530 & 5.36 & 7.73 &13.09 \\
\hline
      \end{tabular}
      }
      \caption{\label{tab:1000fb} As in table~\ref{tab:300fb} above, for $\mathcal{L}=1000$\,fb$^{-1}$.}
\end{table}

\begin{figure}[t!]
\vspace{-0.5cm}
\begin{center}
  \includegraphics*[width=4.9cm]{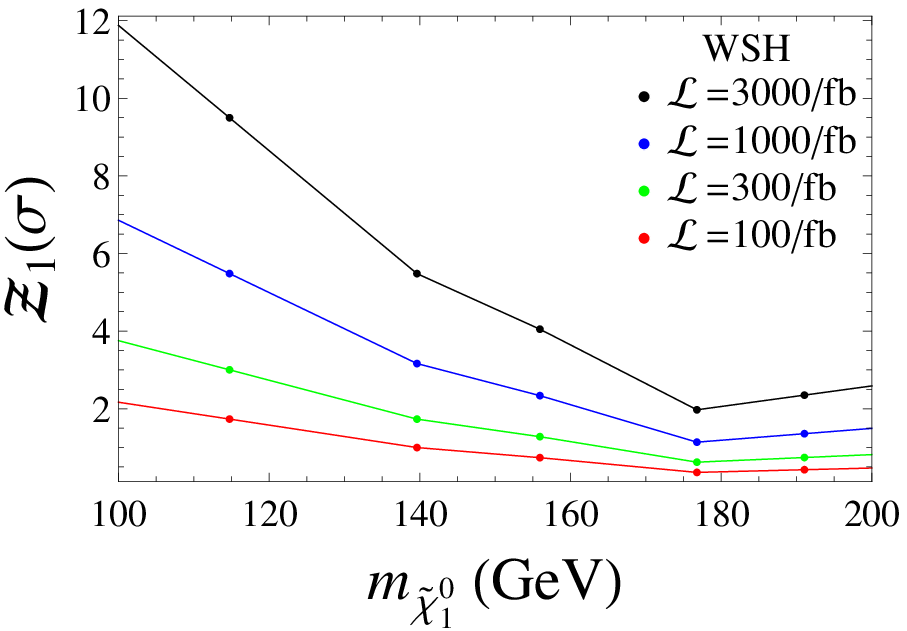}
      \subfloat[]{
        \includegraphics*[width=4.9cm]{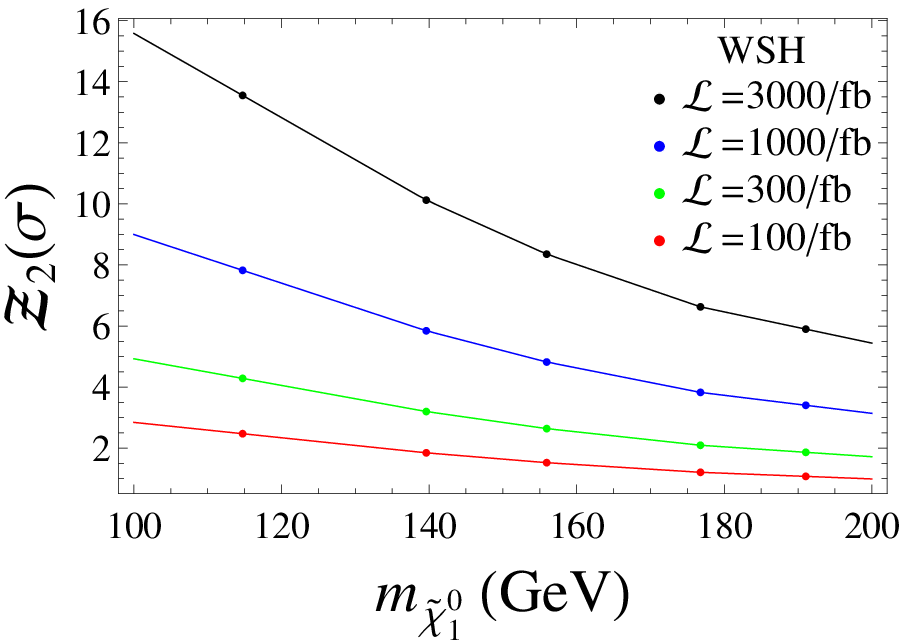}
      }
      \includegraphics*[width=4.9cm]{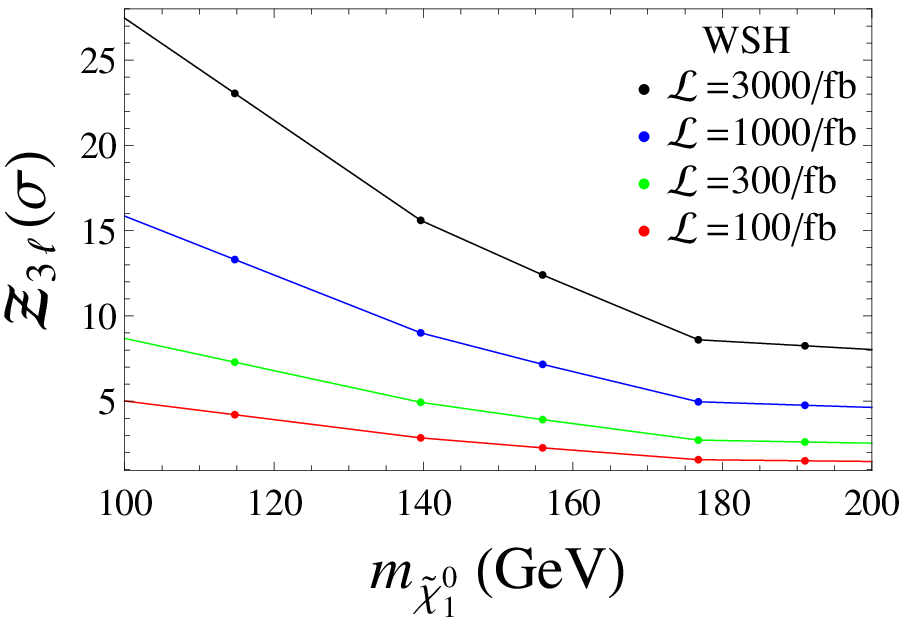} \\
      \includegraphics*[width=4.9cm]{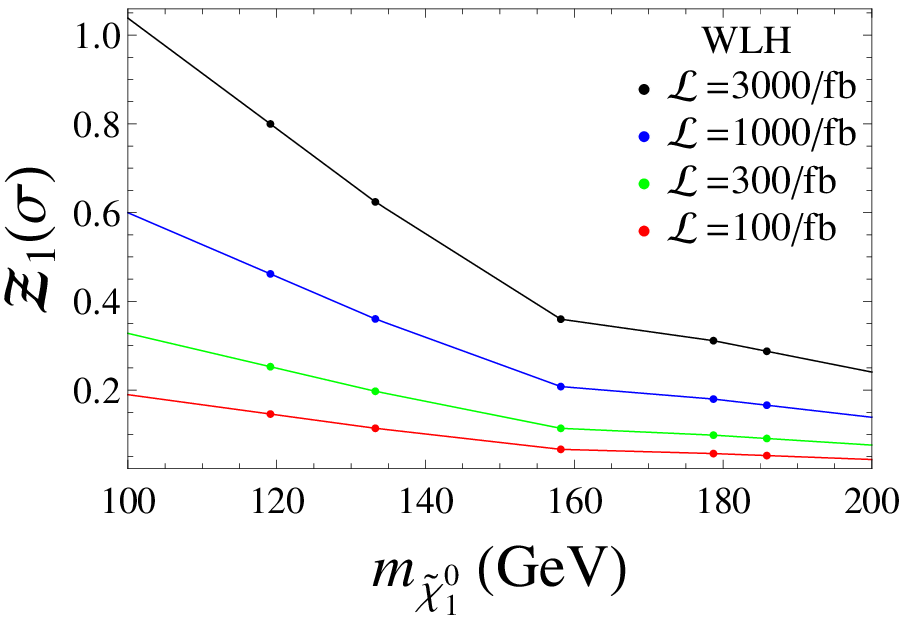}
      \subfloat[]{
      \includegraphics*[width=4.9cm]{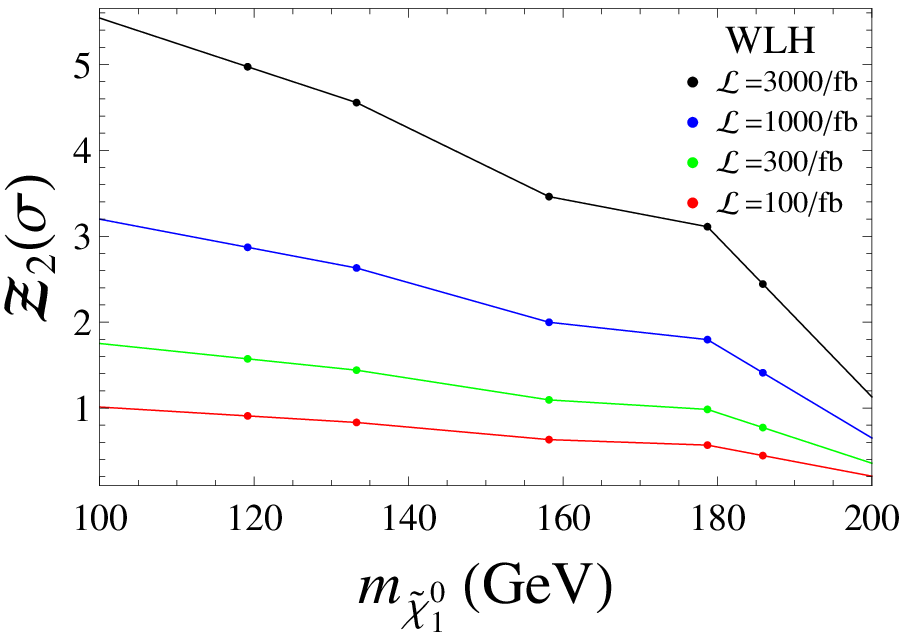}      }
      \includegraphics*[width=4.9cm]{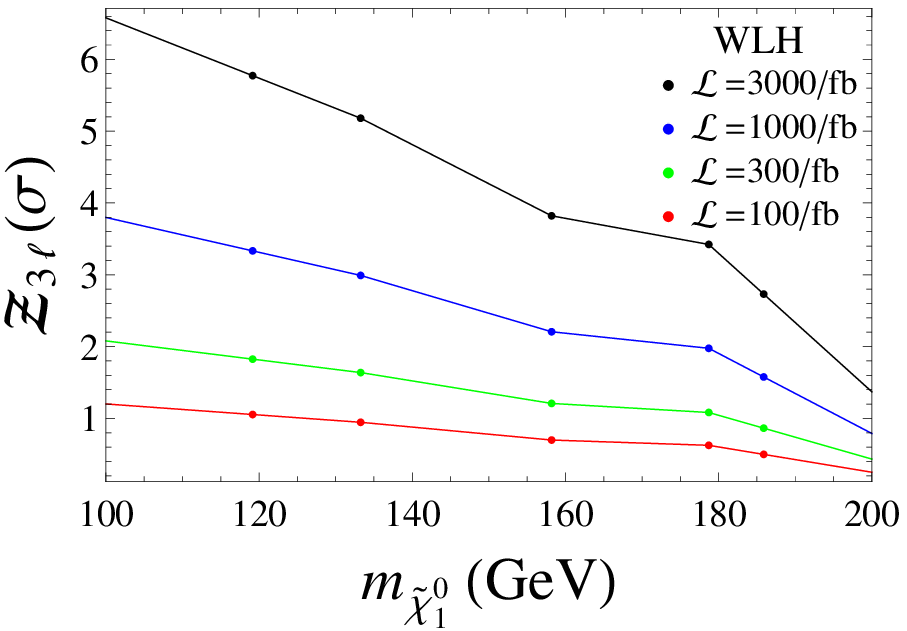}\\
\end{center}
\caption{\label{fig:sigma} Statistical significances obtained for the signals $S_1$ (left column) and $S_2$ (central column) and their combination, $S_{3\ell}$, (right column) for each of the five BPs per scenario. See text for further details.}
\end{figure}

Instead of showing the cut-flow tables and the final significances for all the remaining nine BPs, for brevity, in figure~\ref{fig:sigma} we present our results in the form of contours in the \mneut{1}- $\mathcal{Z}_i$ plane for four different values of $\mathcal{L}$: 100\,fb$^{-1}$ (red), 300\,fb$^{-1}$ (green), 1000\,fb$^{-1}$ (blue) and 3000\,fb$^{-1}$ (black). (We include 3000\,fb$^{-1}$ as the target $\mathcal{L}$ of the so-called High Luminosity LHC (HL-LHC)~\cite{Gianotti:2002xx}). The rows in the figure correspond to the scenarios WSH (a) and WLH (b), and the columns to the signals $S_1$ (left), $S_2$ (centre), and their combination $S_{3\ell}$ (right). Evidently, the shape of a contour is only a crude depiction of the variation in $\mathcal{Z}_i$ with increasing \mneut{1}, owing to the fact that it is obtained by connecting five mutually widely and unevenly spaced BPs (highlighted by the dots on the contours) selected from a random, rather than a continuous, parameter space scan. We note in the figure that, for $\mneut{1}\sim 100$\,GeV in the WSH scenario, the $\mathcal{Z}_{3\ell}$ can reach higher than 4$\sigma$ for $\mathcal{L}=100$\,fb$^{-1}$, thanks mainly to the contribution from $S_2$. For a same-mass DM in the WLH scenario, a combined significance close to 2$\sigma$ may be obtainable with $\mathcal{L}=300$\,fb$^{-1}$, again, with $S_2$ dominating $S_1$ by far. 

\section{\label{sec:concl}Conclusions}

The LHC searches for  supersymmetric DM in the trilepton plus missing transverse momentum channel have by design had maximal sensitivity to a bino-like LSP that emerges in the decays of a pair of the next-to-lightest neutralino and the lightest chargino, both of which are supposedly wino-like. In this study, we have first performed a careful examination of the properties of the wino-like DM and of the accompanying charginos and heavier neutralinos, by defining effective quantities that help establish a holistic and clear overview of the interplay among their masses. Our inferences from the picture that emerged led us to perform a detector-level analysis of some BPs that are consistent with crucial experimental constraints. These include the signal rates of the observed Higgs boson as measured by the LHC, the relic abundance of the universe, the measurements of certain $b$-physics observables, and most importantly, the recent exclusion limits from the XENON1T detector and from the Fermi-LAT and MAGIC experiments combined. 

Through this analysis, we have noted that, while the standard production mode generally results in a relatively poor sensitivity for the DM of our interest, there are regions in the MSSM parameter space where the net yield in the trilepton final state can be substantially enhanced. This is thanks to an alternative channel, the production of the wino-like DM directly in association with the heavier chargino, coming into play. We have demonstrated that, through some optimisation of the kinematical cuts currently employed by the LHC collaborations in the trilepton searches, the complementarity of the two channels can be fully exploited. In particular, we have proposed to drop the $M_T$ cut, the main purpose of which is conventionally to reduce the $W^\pm Z$ background. In the case of the wino-like DM though, the third lepton (coming from the $W^{\pm *}$) is very soft, so that the background essentially contains only two leptons. Hence, instead of this selection, our suggested modified cuts on $\met$ and $M_{\ell^+ \ell^-}$ can be utilised to isolate the combination of the two signals from the SM backgrounds.

The relevance of our analysis for the phenomenology of the MSSM is thus very clear-cut, as a hint of such wino-like DM at the LHC would point towards the following configuration of SUSY:
\begin{enumerate}
\item $\mu < 0$, 
\item an inverted EWino mass hierarchy incompatible with an mSUGRA-inspired SUSY-breaking mechanism, 
\item possibly AMSB dynamics instead, 
\item non-thermal DM production or a multi-component DM,
\end{enumerate}
with, crucially, consistency with the naturalness conditions still plausible, since $|\mu|$ can be sufficiently small.

\acknowledgments

WA would like to thank Ashraf Kassem for useful discussions about the techniques of lepton reconstruction and identification used within the CMS detector at the LHC. The authors thank Korea Institute for Advanced Study for providing computing resources (Linux Cluster System at KIAS Center for Advanced Computation) for this work. This project has received support from the European Union's Horizon 2020 research and innovation programme under the Marie Sk\l{}odowska-Curie grant agreement No. 690575. The work of WA, SK and SMo was partially supported by the H2020-MSCA-RISE-2014 grant No. 645722 (NonMinimalHiggs). SMo is supported in part through the NExT Institute.

%

\bibliography{MSSM_wino-v02.bbl}

\end{document}